\documentclass[apl, aps, two column,showpacs,floatfix]{revtex4-1}
\usepackage{comment}
\usepackage{xcolor}
\usepackage{multirow}
\usepackage{graphics,graphicx,epsfig}
\usepackage[caption=false]{subfig}
\usepackage{amssymb,amsmath}
\usepackage{epstopdf}
\usepackage{amsmath}
\usepackage{bbold}
\usepackage{mathtools}
\usepackage{hyperref}
\hypersetup{colorlinks=true, citecolor = blue, linkcolor=blue, filecolor=magenta, urlcolor=blue}
\begin{document}
\title{Quantum synchronization between two strongly driven YIG spheres mediated via a microwave cavity}

\author{Jatin Ghildiyal}
\email{jatin.20phz0017@iitrpr.ac.in} 
\author{Shubhrangshu Dasgupta}
\author{Asoka Biswas}
\affiliation{Department of Physics, Indian Institute of Technology Ropar, Rupnagar, Punjab 140001, India}

\begin{abstract}

We present a theoretical study of synchronization between two strongly driven magnon modes indirectly coupled via a single-mode microwave cavity. Each magnon mode, hosted in separate Yttrium Iron Garnet spheres, interacts coherently with the cavity field, leading to cavity-mediated nonlinear coupling.  We show, by using input-output formalism, that both classical and quantum synchronization emerge for appropriate choices of coupling, detuning, and driving. We find that thermal noise reduces quantum synchronization, highlighting the importance of low-temperature conditions. This study provides useful insights into tunable magnonic interactions in cavity systems, with possible applications in quantum information processing and hybrid quantum technologies.

\end{abstract}
\maketitle
\section{INTRODUCTION}

In spontaneous synchronization, interacting systems naturally align their dynamical behavior, a trait often observed in nature and society. Synchronization was first reported in a classical context by Huygens in 1665 \cite{huygens1966christiani}, who observed spontaneous phase locking between two pendulum clocks coupled through a common support. This effect, arising from weak mechanical interaction, represents an early example of collective dynamics in coupled oscillators. Such synchronized oscillations can persist despite differences in their natural frequencies \cite{pikovsky2001universal,strogatz2018nonlinear,huygens1966christiani,arenas2008synchronization}. Since then, synchronization has been widely studied in diverse classical systems such as biological networks, electronic circuits, lasers, and mechanical resonators.

While extensively studied in classical contexts, recent research has extended this field into the quantum domain \cite{buvca2022algebraic,e26050415}. Synchronization has been observed or predicted in a variety of quantum systems, ranging from subatomic particles to mechanical resonators \cite{manzano2013avoiding,manzano2013synchronization} and electrodynamic setups \cite{buvca2022algebraic,eneriz2019degree}. These systems often involve multiple interacting modes that exhibit nonlinear dynamics and are subject to strong external drives. For instance, recent experiments demonstrated quantum synchronization of a single trapped ion qubit with an external signal \cite{zhang2023quantum}. 
Quantum synchronization has also been investigated across various other platforms, namely, van der Pol oscillators \cite{giorgi2012quantum}, Josephson junction arrays \cite{wiesenfeld1996synchronization}, spin torque nano-oscillators \cite{kaka2005mutual}, and optomechanical systems \cite{holmes2012synchronization, zhang2012synchronization,heinrich2011collective,PhysRevLett.111.103605,shim2007synchronized}. It has been shown that quantum fluctuations, nonclassical correlations, and phase transitions play critical roles in quantum synchronization phenomena \cite{buvca2022algebraic,eneriz2019degree,PhysRevA.102.032414}. Recently, we have developed a theoretical framework to analyze quantum synchronization between two finite-size spin chains that interact indirectly through a central spin chain \cite{ghildiyal2025quantum}. This configuration results in effective nonlinear coupling, with dynamics influenced by both linear and nonlinear dissipation.

In this paper, we focus on quantum synchronization between two magnon modes. Magnons are quanta of spin-wave excitations arising from the collective precession of localized spins \cite{princep2017full,PhysRevResearch.2.022027,ghasemian2023dissipative,kusminskiy2021cavity}. It has recently been shown that propagating spin waves can mediate phase-tunable synchronization between spin Hall nano-oscillators \cite{kumar2025spin}.  

The ability to couple magnons with cavity modes and qubits has further led to a plethora of work related to quantum information processing in cavity-magnonic systems \cite{YUAN20221}. This development has made it possible to explore quantum synchronization on such a platform. For example, it is shown that magnon squeezing can significantly enhance entanglement between two indirectly coupled magnon modes in a magnomechanical system \cite{harraf2025quantum}. Yang and his coworkers demonstrated asymmetric quantum synchronization between two magnon modes in a two-sublattice antiferromagnet coupled to a cavity \cite{yang2022asymmetric}, highlighting a strong dependence on cavity resonance and input current direction.  We also note that
the magnetostriction in a cavity magnomechanical system enables robust synchronization between two mechanical modes, via nonlinear dispersive coupling \cite{PhysRevResearch.5.043197}. Nonreciprocal quantum synchronization in antiferromagnet-cavity systems has been reported in \cite{yang2021implementationenhancementnonreciprocalquantum}. On the contrary, in our work, we show that magnon-magnon synchronization is indeed possible, just via their common coupling with a single-mode cavity, without requiring any squeezing or magnetostriction. A strong driving field in the presence of an intermode Kerr nonlinear interaction can lead to complete quantum synchronization. We have established our results using detailed time-dependent dynamics and the steady state behaviour of the magnon modes. 

The Yttrium Iron Garnet (YIG) spheres host well-defined magnon modes governed by magnetostatic solutions to the Landau-Lifshitz-Gilbert equation, thanks to their spherical structure. These spheres are often preferred in the studies of magnons due to their low magnetic damping and high coherence, enabling long-lived spin excitations. By placing the YIG spheres inside a high-quality microwave cavity, strong coupling between the magnon modes and the cavity photons can be achieved. This interaction leads to the formation of hybridized states known as cavity magnon polaritons, which are important for quantum technologies involving signal processing, coherent transduction, and nonlinear dynamics \cite{zhang2017observation}. In this paper, we focus on quantum synchronization between two magnon modes of two such spatially separated YIG spheres, placed inside a cavity.

The emergence of synchronization in the present cavity-mediated magnon system originates from the cooperative interplay of four key physical ingredients that are coherent cavity-induced coupling, strong external driving, intrinsic Kerr nonlinearity, and dissipation-assisted stabilization. Although the two magnon modes do not interact directly, their mutual coupling to a common cavity field generates an effective long-range interaction that enables indirect energy and phase exchange. Strong external drives populate the magnon and cavity modes with large coherent amplitudes, pushing the system far from equilibrium and leading to self-sustained oscillations in phase space. The Kerr nonlinearity introduces amplitude-dependent frequency shifts that dynamically compensate detuning mismatches and enable nonlinear frequency entrainment, thereby enabling robust phase locking even for unequal detunings or driving strengths. We emphasize that such a steady-state behavior becomes possible only in the presence of dissipation that plays a constructive role by suppressing unstable trajectories and selecting a unique synchronized attractor, thereby stabilizing the collective dynamics.

We analyze the effects of key parameters, such as coupling strength, detuning, and nonlinear interactions between magnons, and identify the conditions under which stable synchronization emerges. Unlike other relevant works \cite{tyagi2024noise,bittner2025noise}, we have employed the input-output formalism, instead of the master equation technique, to analyze the effect of decoherence. This formalism naturally incorporates dissipation and quantum noise through Heisenberg–Langevin equations. It provides a computational advantage over master equation approach while working with an infinite dimensional Hilbert space, multimode or non-linear systems. Additionally, the input–output framework directly connects intracavity dynamics to measurable output fields.

The paper proceeds as follows. In Section II, we present the theoretical model, including the mean-field approximation and the corresponding dynamical equations. In Section III, we analyze the main results on synchronization, focusing on the emergence of limit cycles and relevant synchronization measures. We will discuss the experimental feasibility of our model. Finally, we conclude the paper in Section IV.

\section{MODEL AND EQUATION OF MOTION}
We consider two macroscopic YIG spheres, trapped inside a single-mode microwave cavity. Each sphere hosts many magnon modes. The lowest-order mode, known as the Kittel mode, corresponds to the collective spin ground state, featuring uniform spin precession across the sample \cite{princep2017full}. The higher-order modes exhibit non-uniform spatial distributions with specific angular and radial profiles. Here, we consider the YIG spheres in their respective Kittel mode. The cavity mode with the annihilation operator $a$ interacts coherently with these magnon modes with respective annihilation operators $m_1$ and $m_2$. This setup allows us to create an indirect coupling between the magnon modes through the cavity. 

\begin{figure}[ht]
\includegraphics[height=4cm,width=6cm]{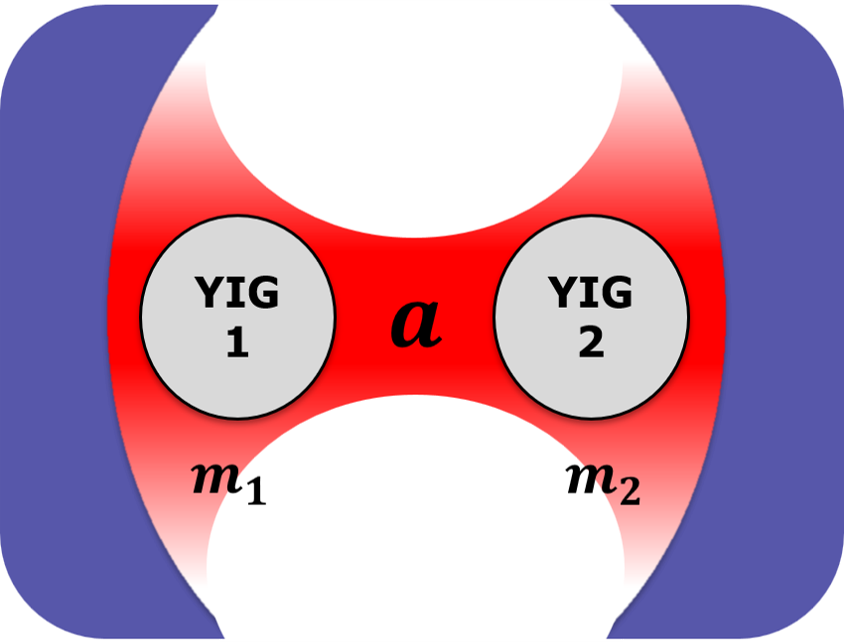}
    \caption{Schematic representation of two YIG (Yttrium Iron Garnet) spheres, ${m}_1$ and ${m}_2$, coupled via a common microwave cavity mode, ${a}$.}
    \label{fig:YIG_coupling}
\end{figure}

The cavity is driven by an electromagnetic field of frequency $\omega_L$.  Typically, the magnetic component of this driving field couples with the magnon mode of the macroscopic ensemble inside the sphere, via Zeeman interaction. The interaction Hamiltonian between the magnon modes and the cavity can then be written as
\cite{PhysRevB.94.224410, PhysRevLett.104.077202}
\begin{equation}
H_I = \sum_{i=1}^2 g_i \left({a}^{\dagger} {m}_i +{m}^{\dagger}_i {a} \right)\;,  \end{equation} 
where $g_i$ is the coupling constant for the interaction between the cavity mode and the $i$th magnon mode. 

The unperturbed Hamiltonians $H_c$ and $H_m$ of the cavity and magnon modes, respectively, are given by
\begin{eqnarray}
H_c&=&\omega_c{a}^{\dagger} {a} 
+ \Omega_{c} \left({a}^{\dagger} e^{-\iota \omega_Lt} + {a} e^{\iota \omega_L t} \right) \;, \label{eq2}\\
H_m&=& \sum_{i=1}^2\left[\omega_{m_i} {m}^{\dagger}_i {m}_i 
+ K_i ({m}^{\dagger}_i {m}_i)^2   \right.\nonumber\\
&&\left.+ \Omega_{i} \left({m}^{\dagger}_i e^{-\iota \omega_{d}^{(i)} t} + {m}_i e^{\iota \omega_{d}^{(i)} t} \right)\right]\;. 
\label{eq3}
\end{eqnarray}
Here $\omega_c$ is the cavity resonance frequency and $\omega_{m_i}$ is the frequency of the $i$th magnon mode. The $\Omega_c$ and $\Omega_i$ denote the Rabi frequencies, describing the amplitudes of the driving fields with respective frequencies $\omega_L$ and $\omega_d^{(i)}$, respectively. The term containing the Kerr coefficient $K_i$ arises from the magnetocrystalline anisotropy in YIG, which introduces intensity-dependent frequency shifts proportional to the magnon population $\langle m_i^\dagger m_i\rangle$ \cite{PhysRevB.94.224410,ganthya2022bistability}. This term leads to a nonlinear phase-inducing evolution \cite{PhysRevB.94.224410, PhysRevResearch.1.023021} of the magnons.

The Hamiltonian (\ref{eq2}) contains the terms for cavity self-energy and the cavity driving, while Eq. (\ref{eq3}) is a sum of magnon self-energy, nonlinear energy shift, and magnon driving term, respectively. In the rotating frame with respect to the driving frequencies $\omega_L$ and $\omega_d^{(i)}$, the explicit time dependence associated with the external drives is removed under the assumption of near-resonant driving and rotating-wave approximation. This enables to remove the rapidly oscillating terms as their time-average vanishes over many cycles of oscillations. This leads to the following time-independent Hamiltonian:
\begin{eqnarray}
H & =& \Delta_c{a}^{\dagger} {a}+\Omega_{c}\left({a}^{\dagger} +{a}\right) 
+ \sum_{i=1}^2 \left\{ \Delta_i {m}^{\dagger}_i {m}_i + K_i ({m}^{\dagger}_i {m}_i)^2\right.\nonumber \\
& & + \left. \Omega_{i}\left({m}^{\dagger}_i+{m}_i\right)+g_i\left({a}^{\dagger} {m}_i+{m}^{\dagger}_i {a}\right) 
 \right\}\;,\label{eq4}
\end{eqnarray}  
where $\Delta_c=\omega_c-\omega_L$ and $\Delta_i=\omega_{m_i}-\omega_d^{(i)}$ are the detunings of the cavity mode and the $i$th magnon mode, respectively, from the corresponding driving field.  

Incorporating the interaction picture with respect to the energy terms containing $a^\dagger a$ and $m_i^\dagger m_i$ and using the Baker–Campbell–Hausdorff formalism, the modified Hamiltonian $H^{\prime}(t)$ can be expressed as $(\iota = \sqrt{-1})$
\begin{eqnarray}
H^{\prime}(t) &=&\sum_{i=1}^2 \left[g_ia^{\dagger}e^{\iota t(B_i-C_i)}m_i + \Omega_i m_i^{\dagger}e^{\iota t(D_i+C_i)}\right]\nonumber\\
& &+\Omega_c a^{\dagger}e^{\iota 
 \Delta_c t}+ H.c.\;,
\end{eqnarray}
where
\begin{eqnarray}
B_i &=& \Delta_c-\Delta_i\;,
\nonumber\\
C_i &=& K_i(m_i^{\dagger}m_i+m_im_i^{\dagger})\;,
\\
D_i &=& \Delta_i\;.
\nonumber
\end{eqnarray}

Note that the above Hamiltonian is inherently nonlinear. When the YIG sphere is driven, the frequency of the magnon shifts due to the Kerr nonlinearity. To study the two-magnons synchronization, we first obtain the Langevin's equations for the magnon modes $m_1$ and $m_2$ and the cavity mode $a$, given by 
\begin{equation}\label{b1eqn}
\begin{aligned}
\dot{m}_i &= -\frac{\gamma_i}{2} m_i +\sqrt{\gamma_i}m_{\text {in}}^{(i)}+ \iota \bigg[ g_i a^{\dagger} A_i e^{\iota t (B_i -  C_i)} m_i^2 \\
& \quad- \Omega_i e^{\iota t (D_i + C_i)}- \Omega_i m_i^{\dagger} A_i e^{\iota t (D_i +  C_i)} \\
& \quad - g_i e^{-\iota t (B_i - C_i)} a - g_i m_i^{\dagger} A_i e^{-\iota t (B_i - C_i)} m_i a \\
&\quad + \Omega_i A_i e^{-\iota t (D_i + C_i)} m_i^2 \bigg] 
\end{aligned}
\end{equation}
\begin{equation}\label{b2eqn}
\begin{aligned}
\dot{a} &= -\frac{\gamma_c}{2} a +\sqrt{\gamma_c} a_{\mathrm{in}} - \iota \Omega_c e^{\iota\Delta_c t} - \iota \sum_{i=1}^2 g_i e^{\iota t (B_i- C_i)} m_i.
\end{aligned}
\end{equation}
where $A_i=2\iota K_i t$ ($i\in 1,2$) is linearly proportional to the time $t$.

In these equations, $\gamma_i$ and $\gamma_{c}$ represent the linear dissipation rates of magnons and cavity, respectively. The input noise operators ${a}_{\text{in}}$ and ${m}_{\text{in}}^{(i)}$ satisfy the following two-time correlation functions:
\begin{equation}
\left\langle a_{\text{in}}(t) a_{\text{in}}^{\dagger}(t') \right\rangle =  \delta(t - t') \;\;,\;\;
\left\langle a_{\text{in}}^{\dagger}(t) a_{\text{in}}(t') \right\rangle = 0\;.
\end{equation}
and
\begin{equation}
\begin{aligned}
\left\langle m_{\text{in}}^{(i)}(t) m_{\text{in}}^{(i)\dagger}(t') \right\rangle &=  \left(\bar{n}_{m} + 1 \right) \delta(t - t')\;, \\
\left\langle m_{\text{in}}^{(i)\dagger}(t) m_{\text{in}}^{(i)}(t') \right\rangle &=  \bar{n}_{m} \delta(t - t')\;.
\end{aligned}
\end{equation}
Here $\bar{n}_m$ is the average phonon number of the thermal bath, common to both magnon modes.

\begin{figure*}

     \subfloat[\label{pqa}]{%
		\includegraphics[height=4.7cm,width=0.34\linewidth]{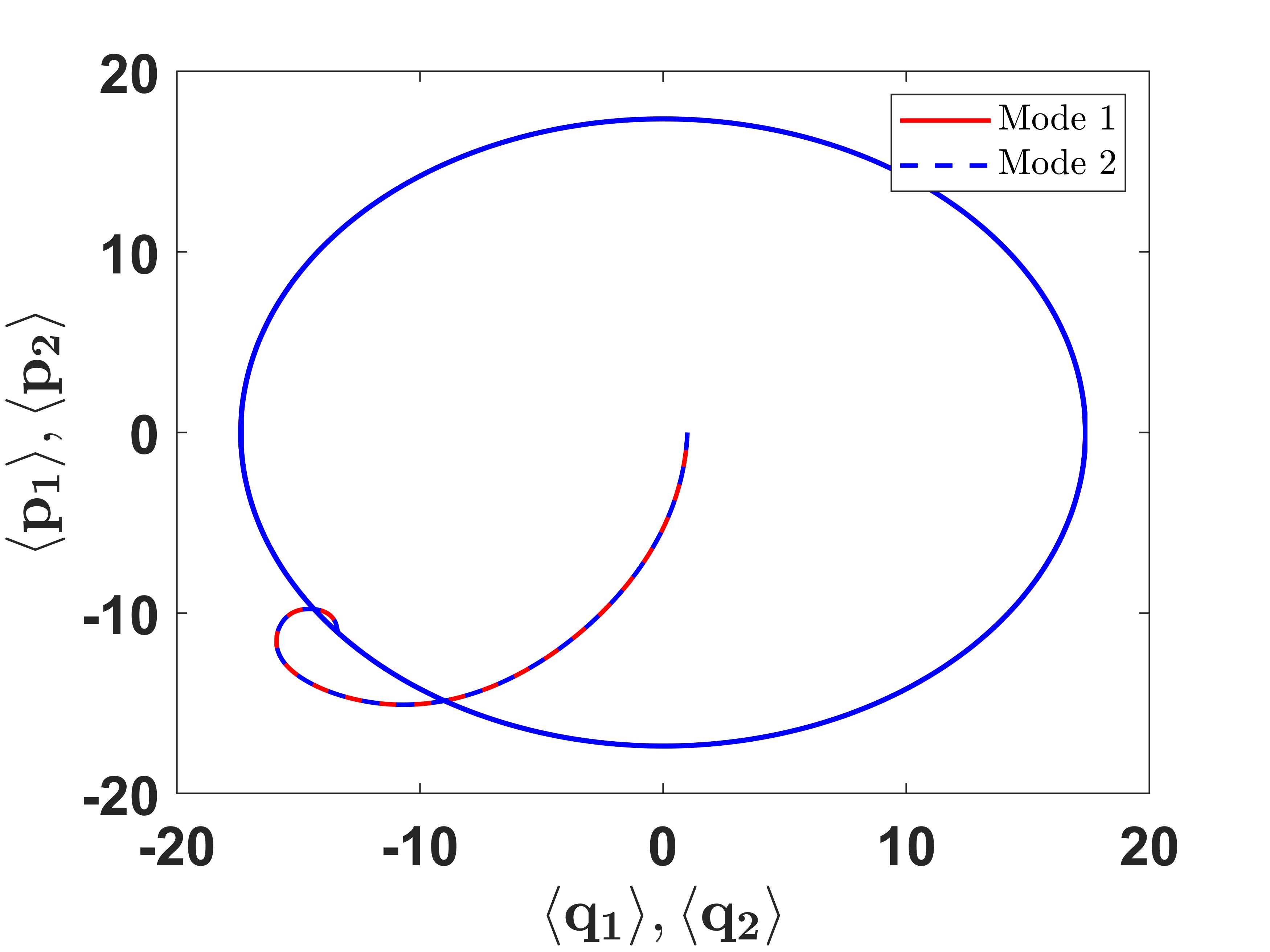}%
	 }
      \subfloat[\label{q1q2a}]{%
		\includegraphics[height=4.7cm,width=0.34\linewidth]{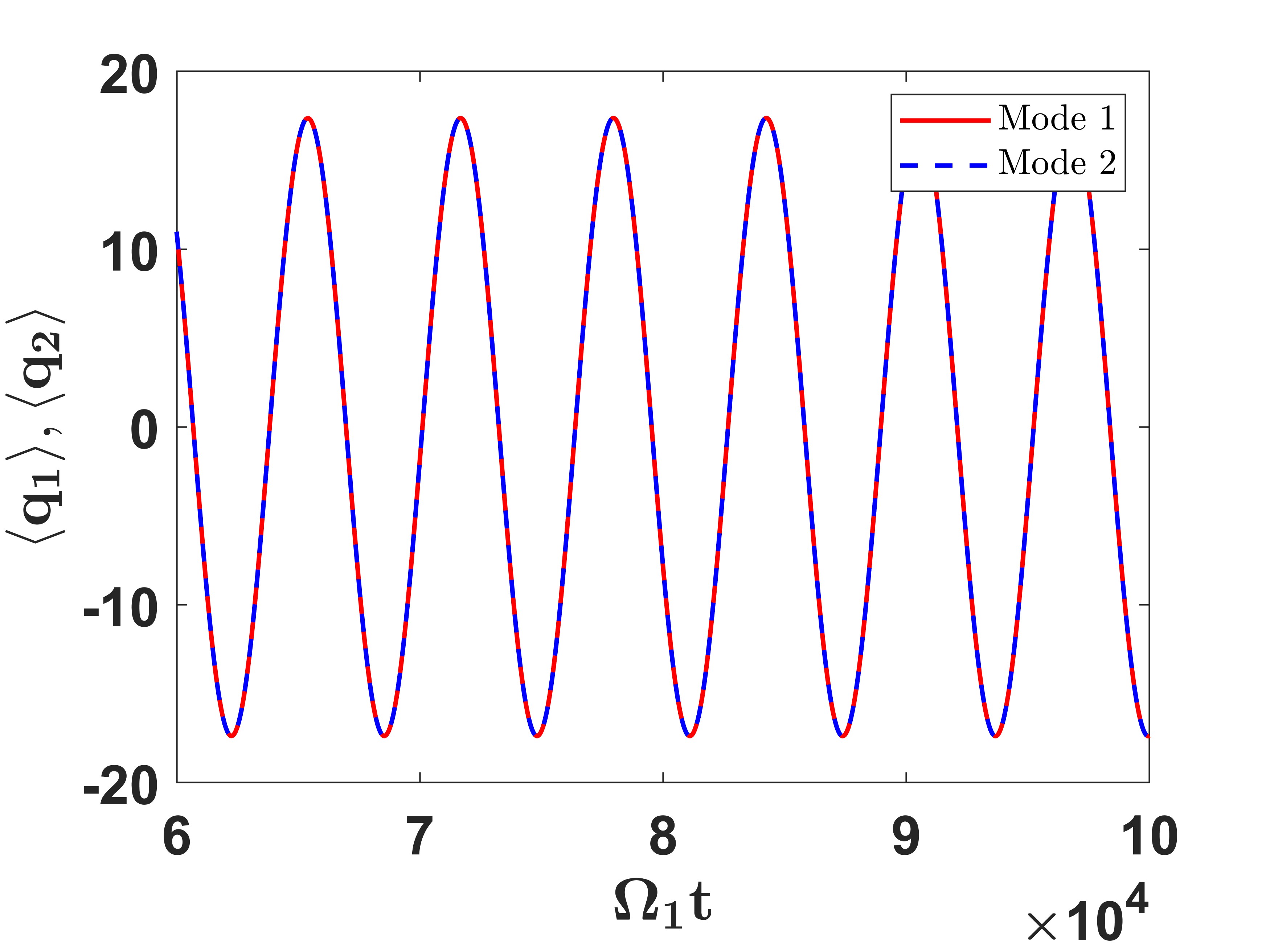}%
	 }
        \subfloat[\label{p1p2a}]{%
		\includegraphics[height=4.7cm,width=0.34\linewidth]{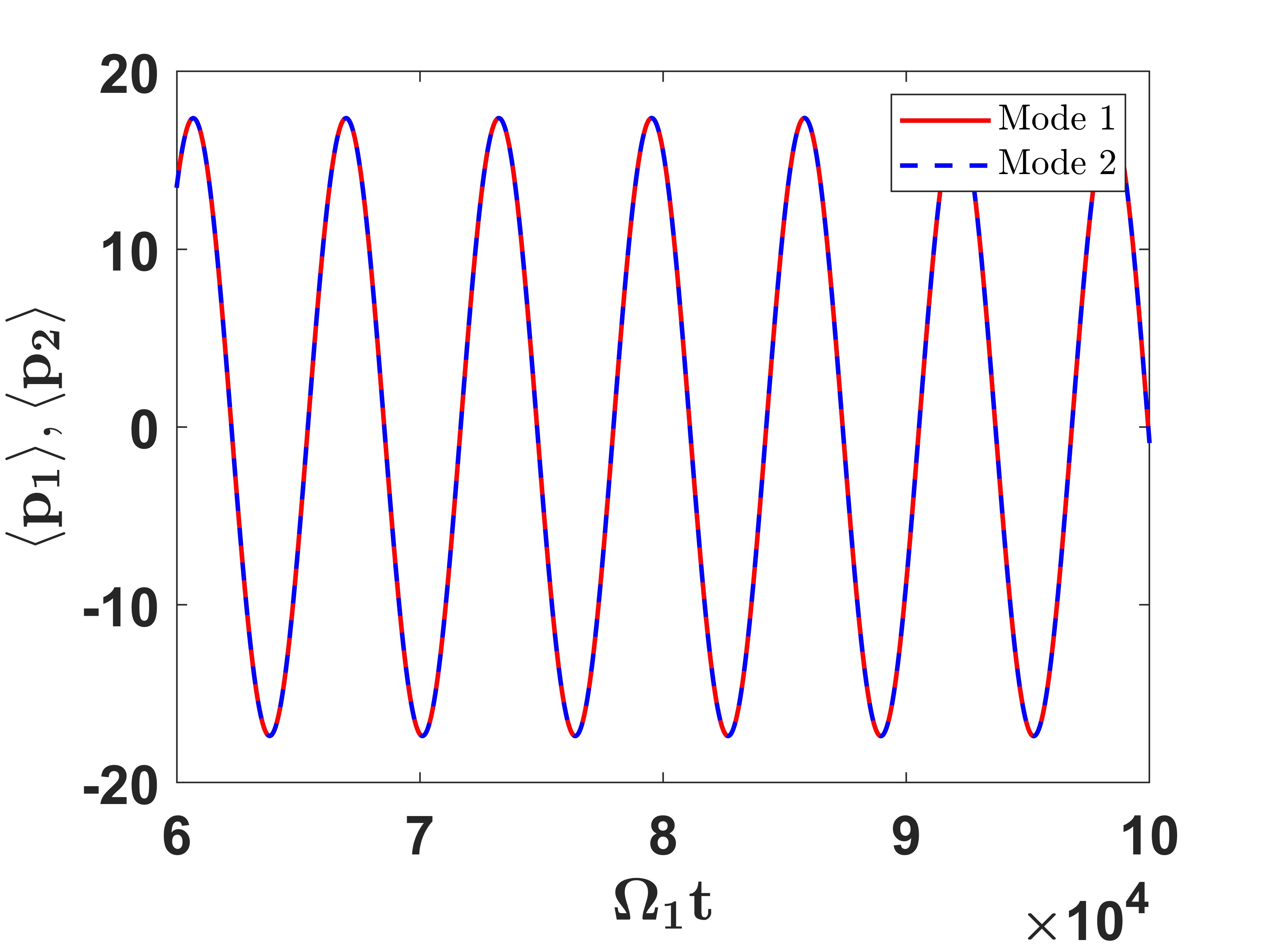}
	}\\

         \subfloat[\label{pqb}]{%
		\includegraphics[height=4.7cm,width=0.34\linewidth]{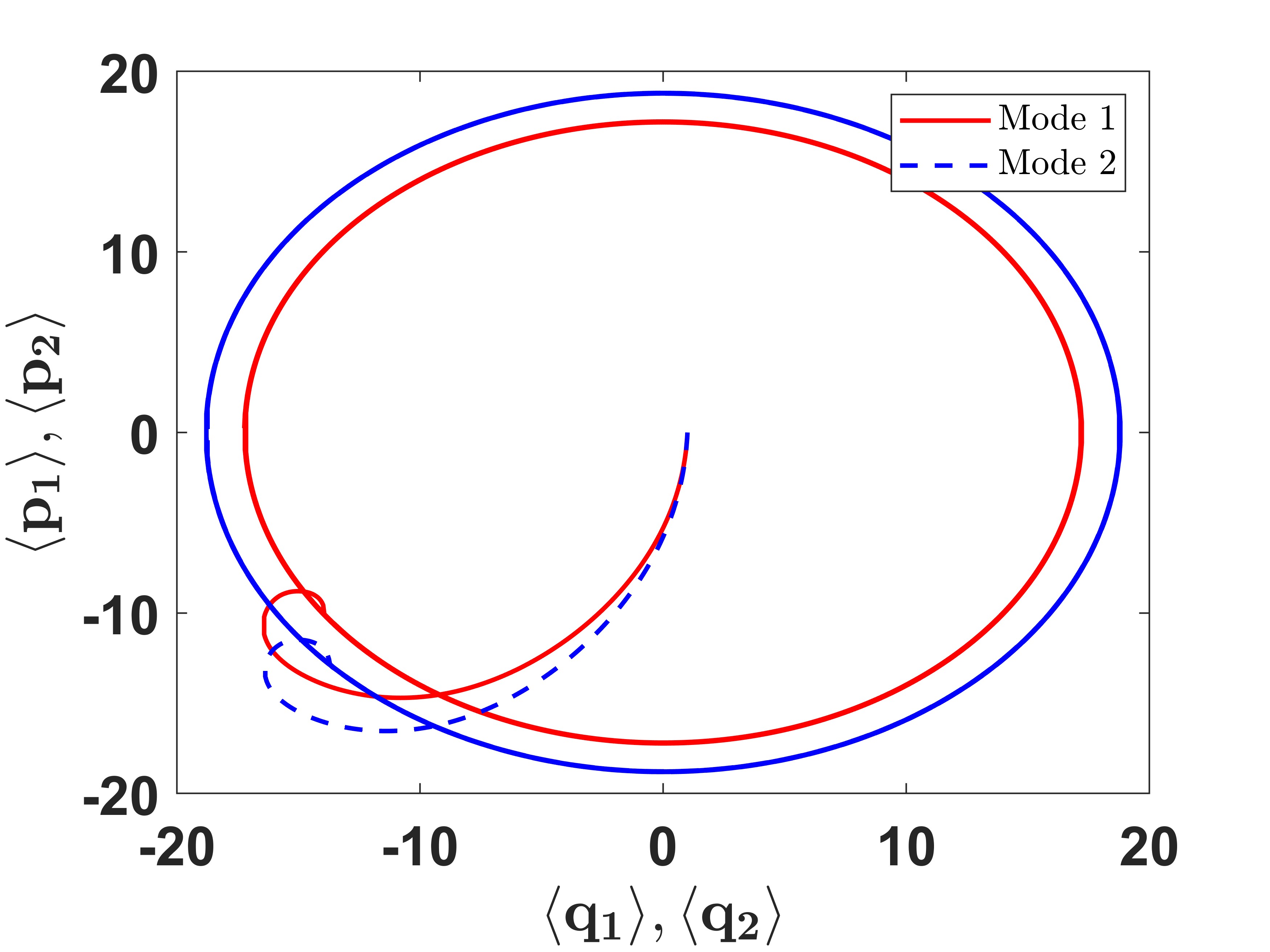}%
	 }
      \subfloat[\label{q1q2b}]{%
		\includegraphics[height=4.7cm,width=0.34\linewidth]{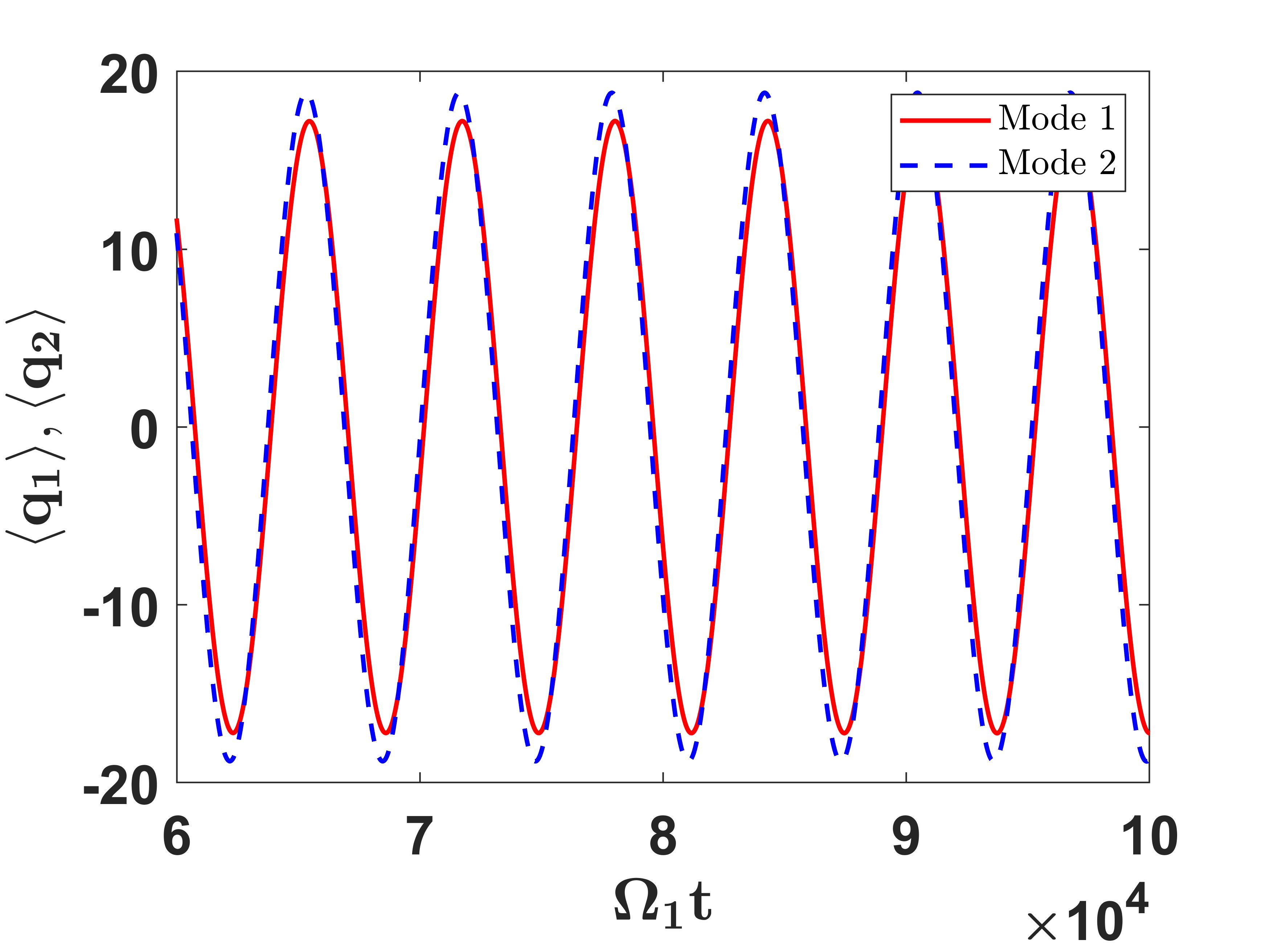}%
	 }
        \subfloat[\label{p1p2b}]{%
		\includegraphics[height=4.7cm,width=0.34\linewidth]{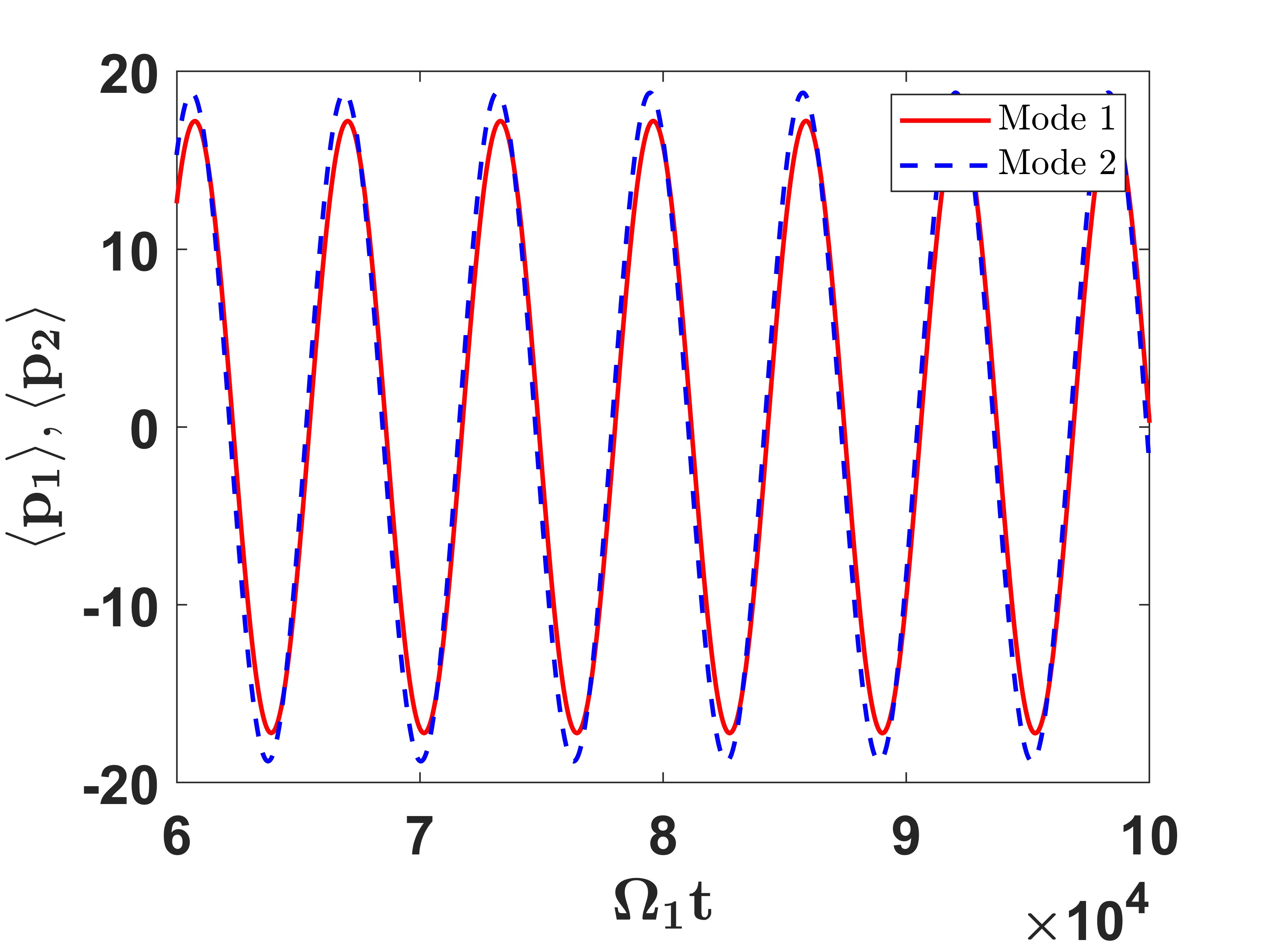}
	}\\
        \subfloat[\label{PQC}]{%
		\includegraphics[height=4.7cm,width=0.34\linewidth]{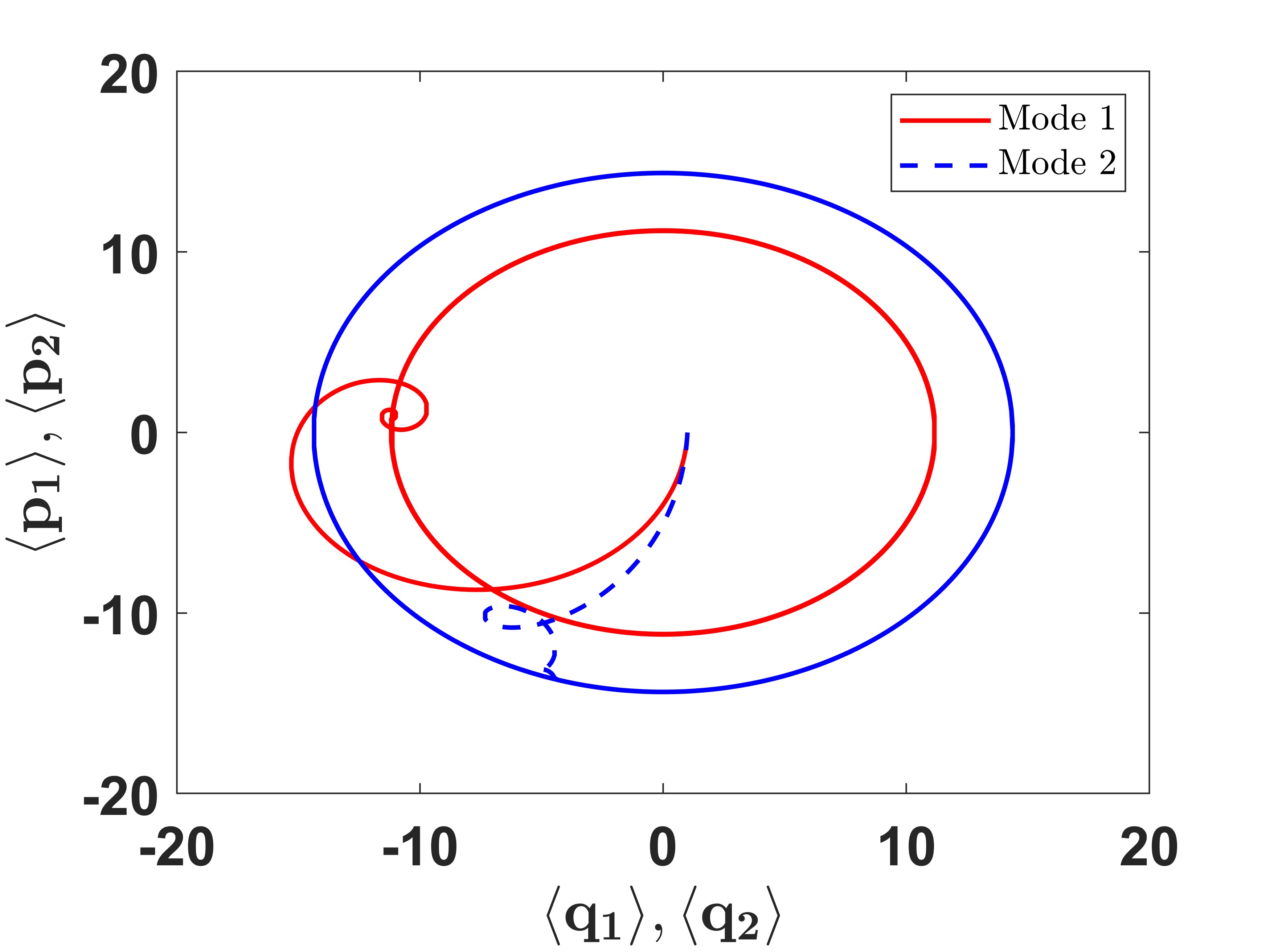}%
	 }
    \subfloat[\label{Q1Q2C}]{%
		\includegraphics[height=4.7cm,width=0.34\linewidth]{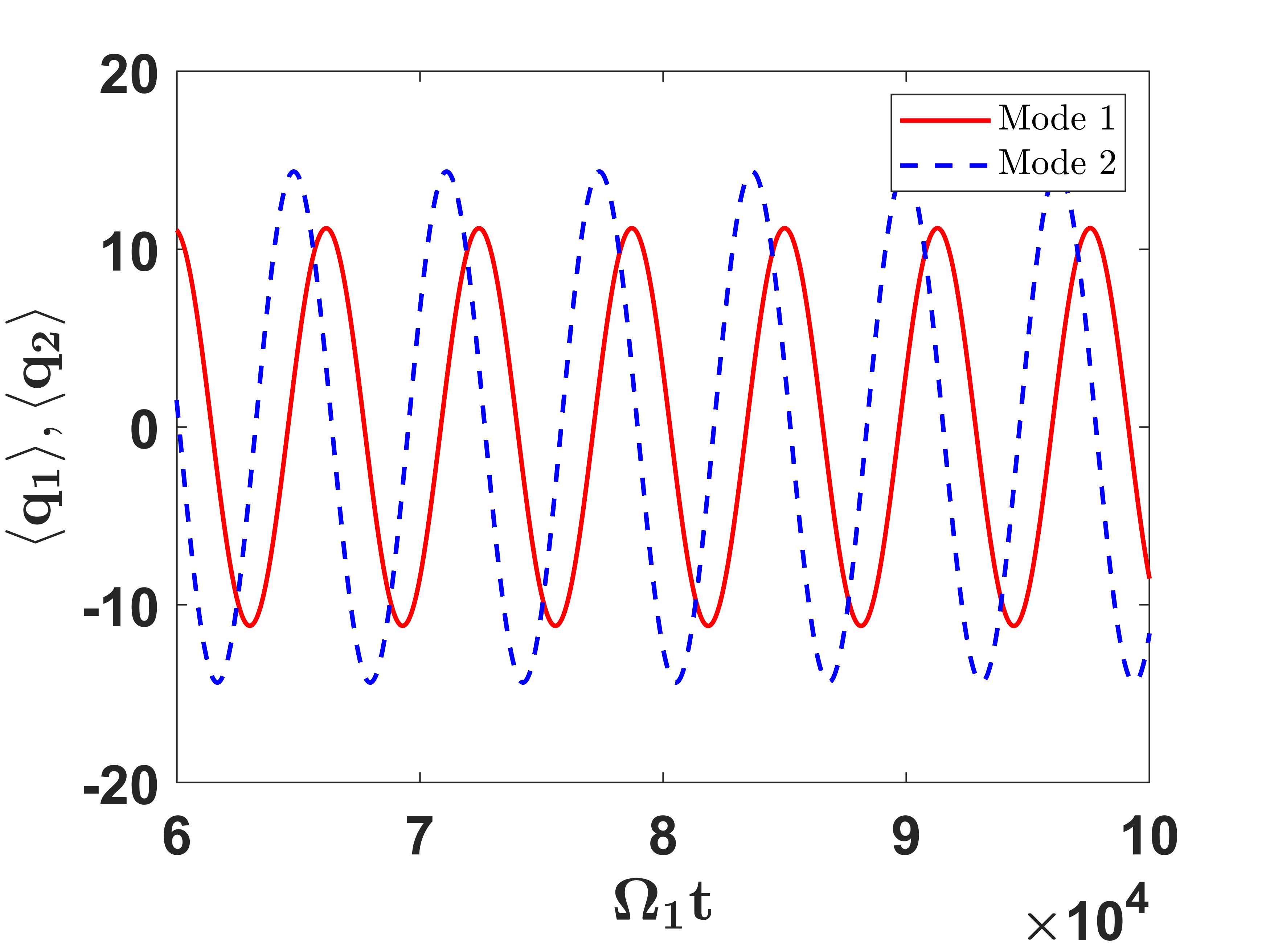}%
	 }
    \subfloat[\label{P1P2C}]{%
		\includegraphics[height=4.7cm,width=0.34\linewidth]{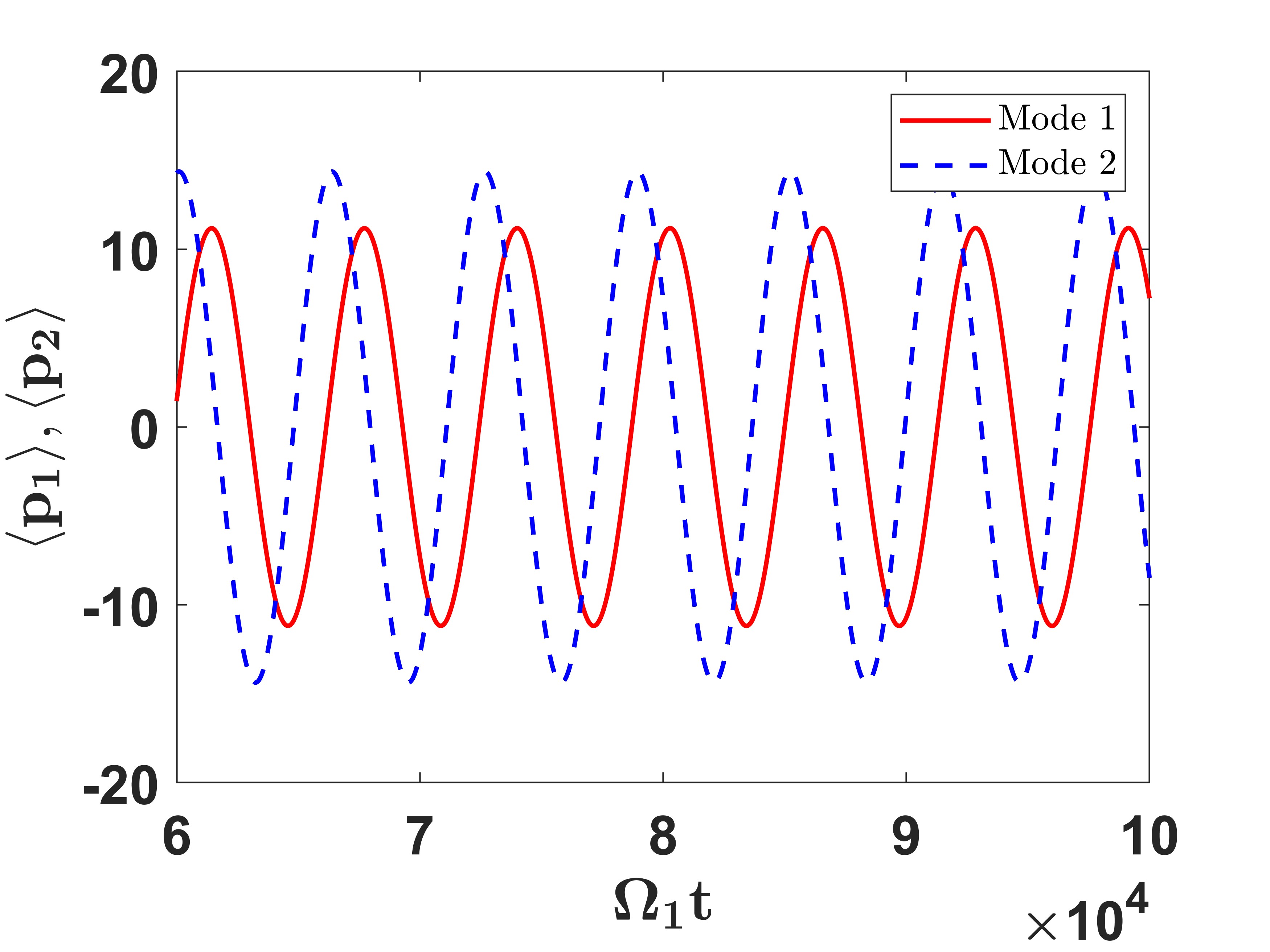}%
	 }
     \\

\caption{Limit-cycle trajectories in the $\langle{q_{1}\rangle} \leftrightharpoons \langle{p_{1}\rangle}$ (red) and $\langle{q_{2}\rangle} \leftrightharpoons \langle{p_{2}\rangle}$ (blue) spaces (a, d, g), and variation of the mean values $\langle{q}_{1}\rangle$ (red) and $\langle{q}_{2}\rangle$ (blue) (b, e, h), along with mean values $\langle{p}_{1}\rangle$ (red) and $\langle{p}_{2}\rangle$ (blue) (c, f, i). The chosen parameters are: (a--c) $\Omega_{1} = 1$, $\Omega_{2} = 1.00001$, $g_{1} = g_{2} = 0.1$. (d--f) $\Omega_{1} = 1$, $\Omega_{2} = 1.1$, $g_{1} = g_{2} = 0.1$. (g--i) $\Omega_{1} = 1$, $\Omega_{2} = 1.00001$, $g_{1} = 0.2$, $g_{2} = 0.1$. All other parameters are identical across cases: $\Omega_{c} = 1$, $\Delta_{1} = \Delta_{2} = 0.001$, $\Delta_{c} = -0.2$, $K_{1} = K_{2} = 10^{-10}$, and $\gamma_{1} = \gamma_{2} = \gamma_{c} = 0.1$. All parameters are normalized with respect to $\Omega_{1}$ with the initial conditions $(\langle{q}_1\rangle,\langle{p}_1\rangle)=(1,0) =(\langle{q}_2\rangle,\langle{p}_2\rangle)$ [i.e., $\alpha_i=1/\sqrt{2}$ well satisfying the mean-field approximation]. The limit cycle trajectories are shown till $\Omega_1t =10^5$.}
\label{limitcycle}
\end{figure*}

\begin{figure*}
       
       \subfloat[\label{PQD}]{%
		\includegraphics[height=4.7cm,width=0.34\linewidth]{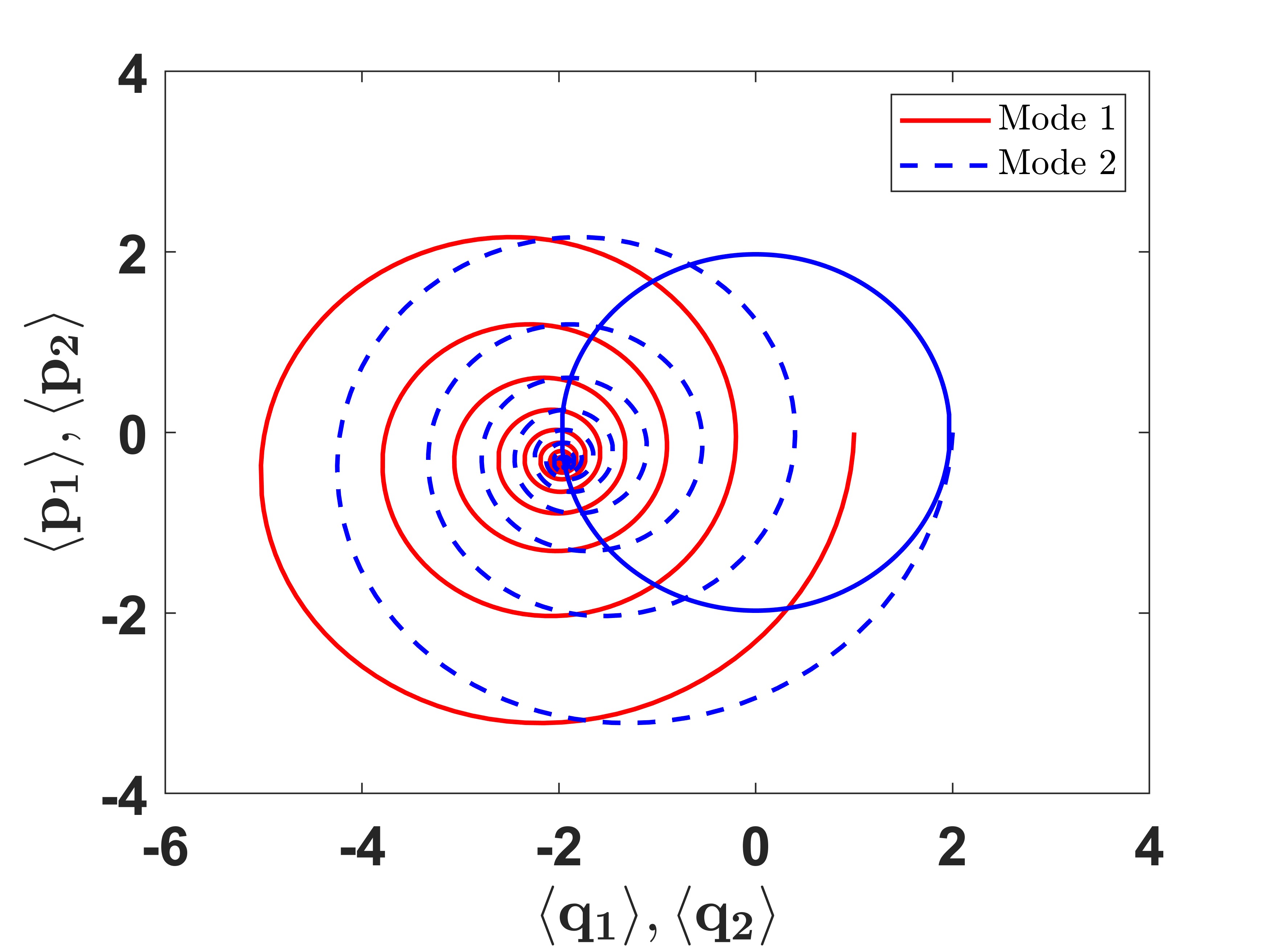}%
	 }
      \subfloat[\label{Q1Q2D}]{%
		\includegraphics[height=4.7cm,width=0.34\linewidth]{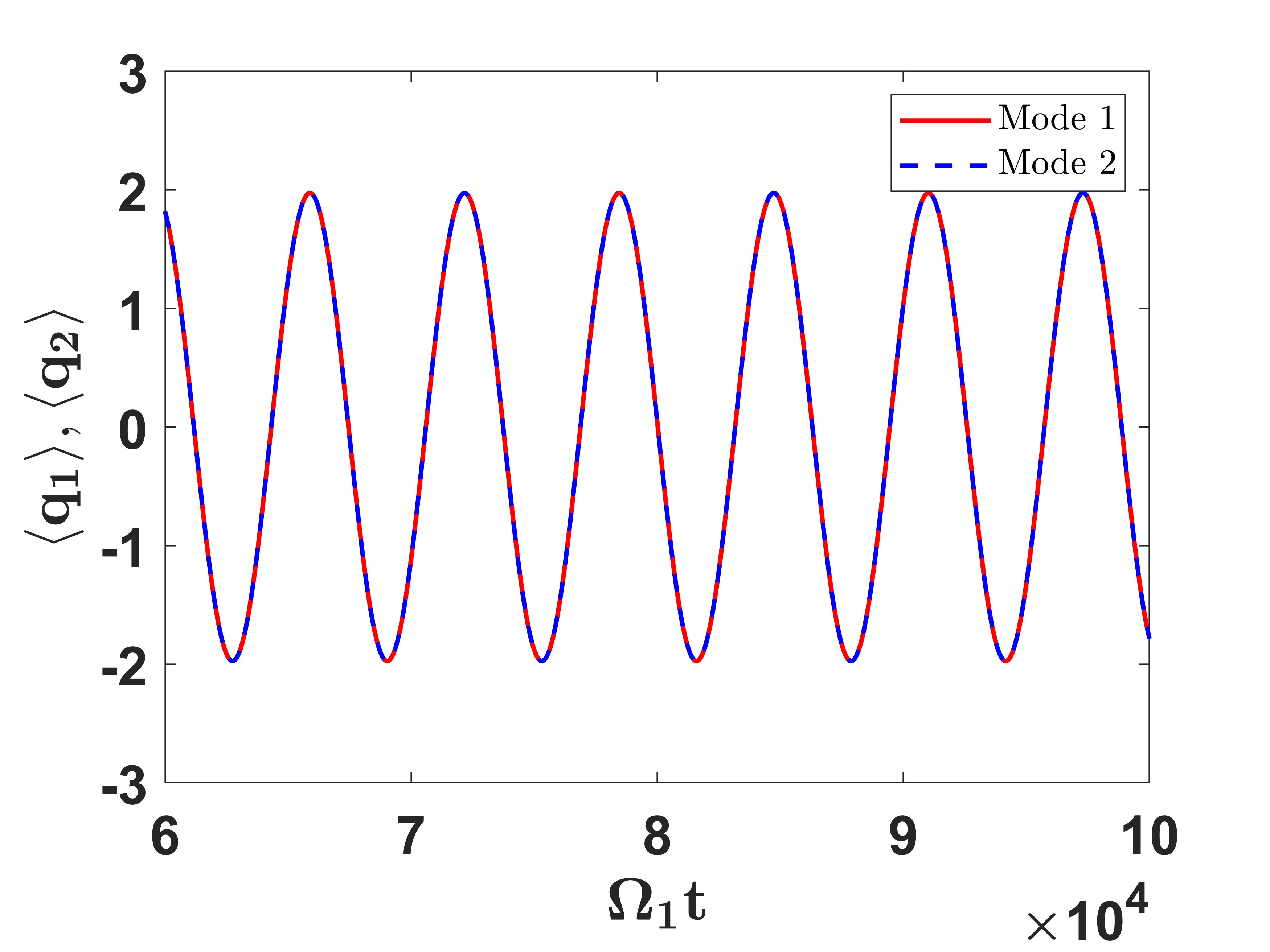}%
	 }
        \subfloat[\label{P1P2D}]{%
		\includegraphics[height=4.7cm,width=0.34\linewidth]{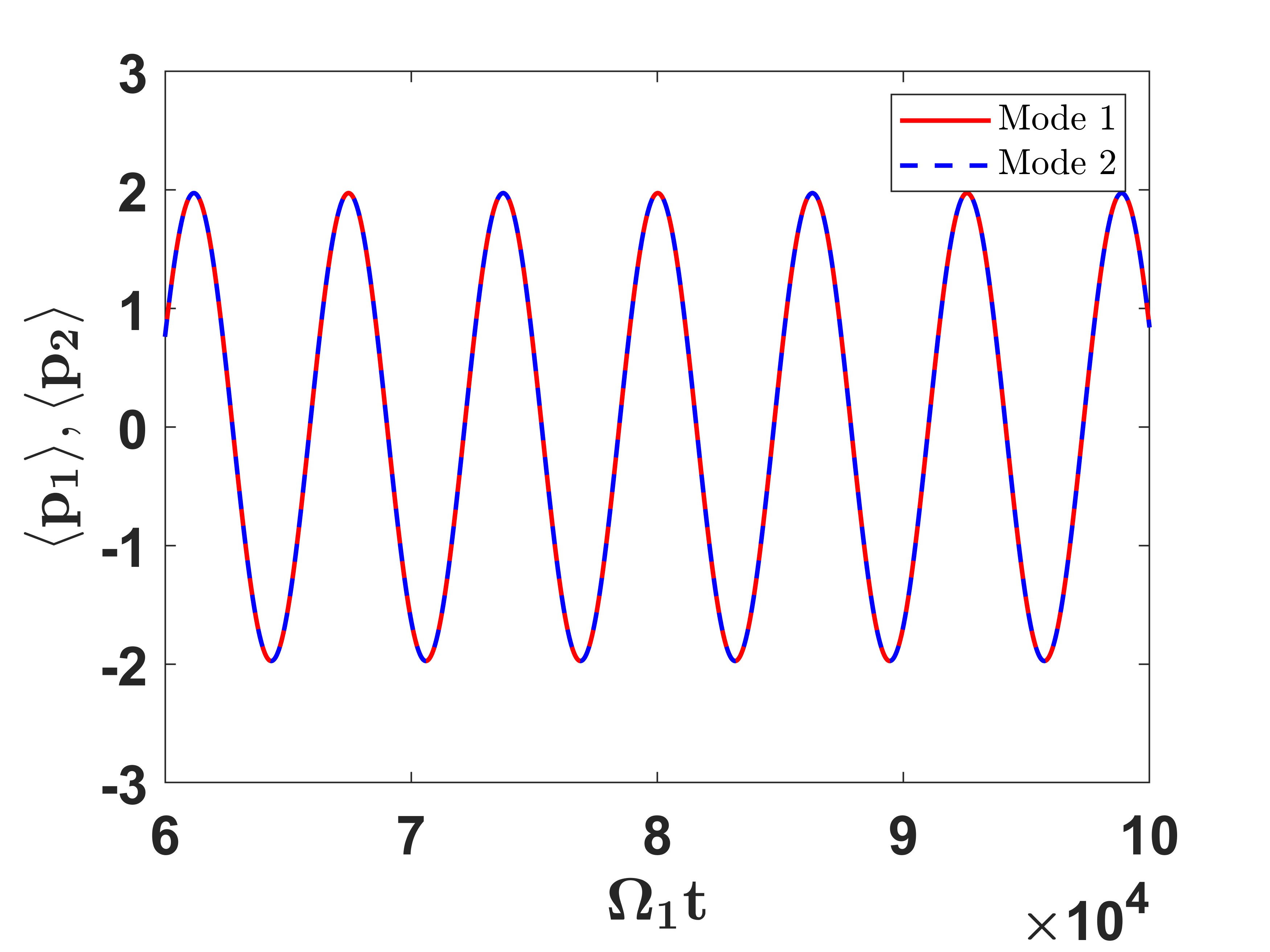}
	}\\

         \subfloat[\label{PQE}]{%
		\includegraphics[height=4.7cm,width=0.34\linewidth]{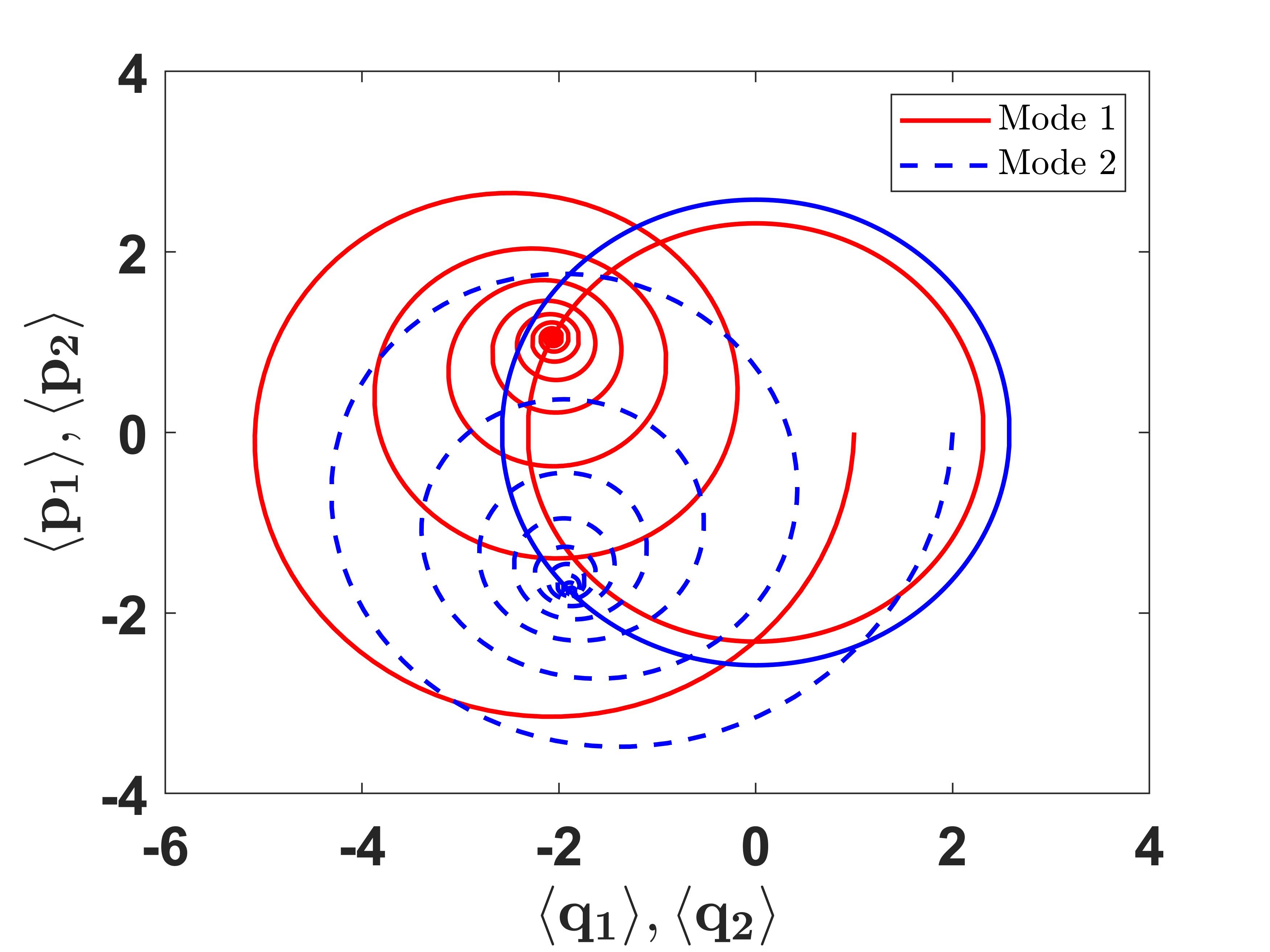}%
	 }
      \subfloat[\label{Q1Q2E}]{%
		\includegraphics[height=4.7cm,width=0.34\linewidth]{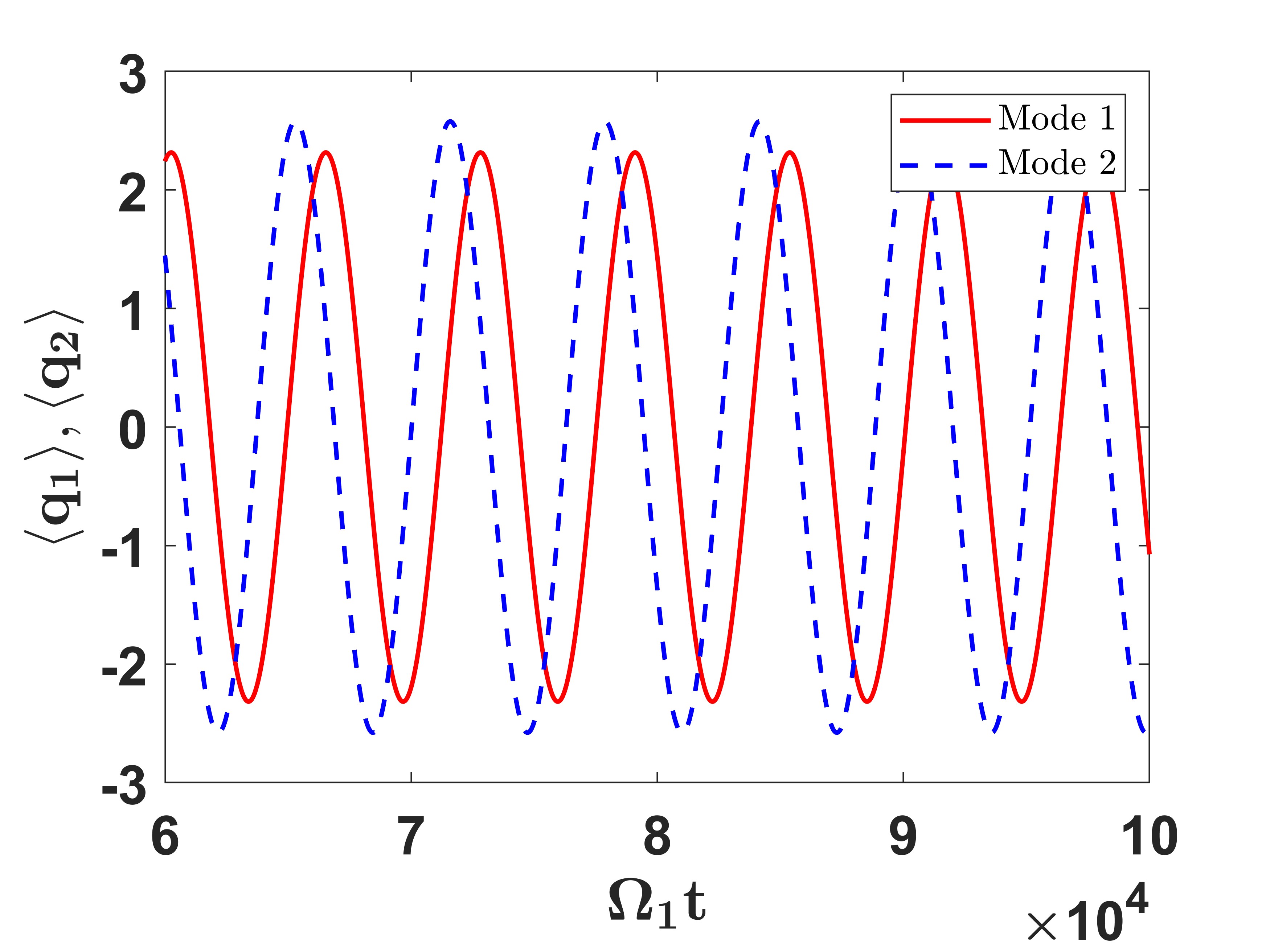}%
	 }
        \subfloat[\label{P1P2E}]{%
		\includegraphics[height=4.7cm,width=0.34\linewidth]{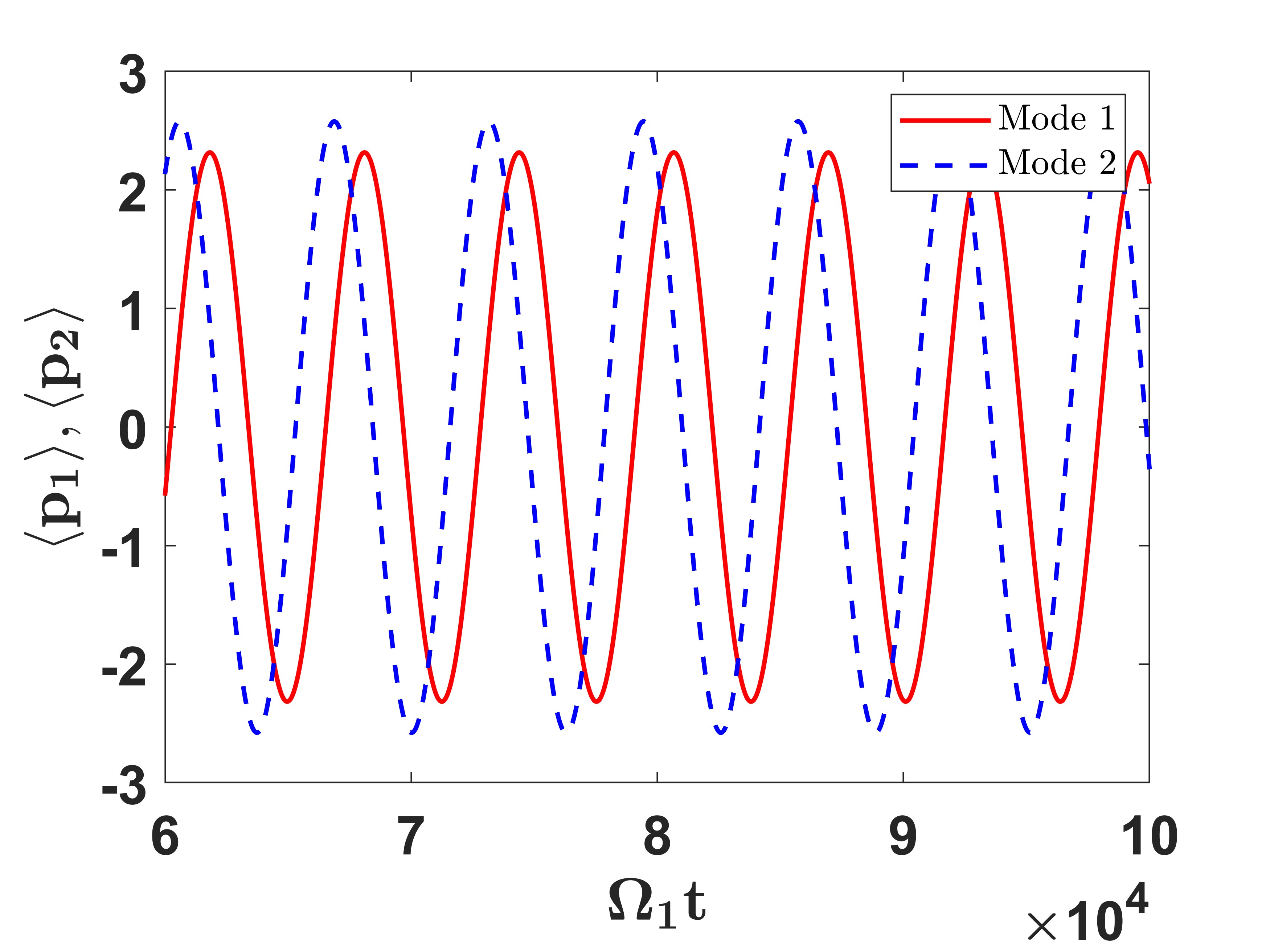}
	}\\

             \subfloat[\label{PQF}]{%
		\includegraphics[height=4.7cm,width=0.34\linewidth]{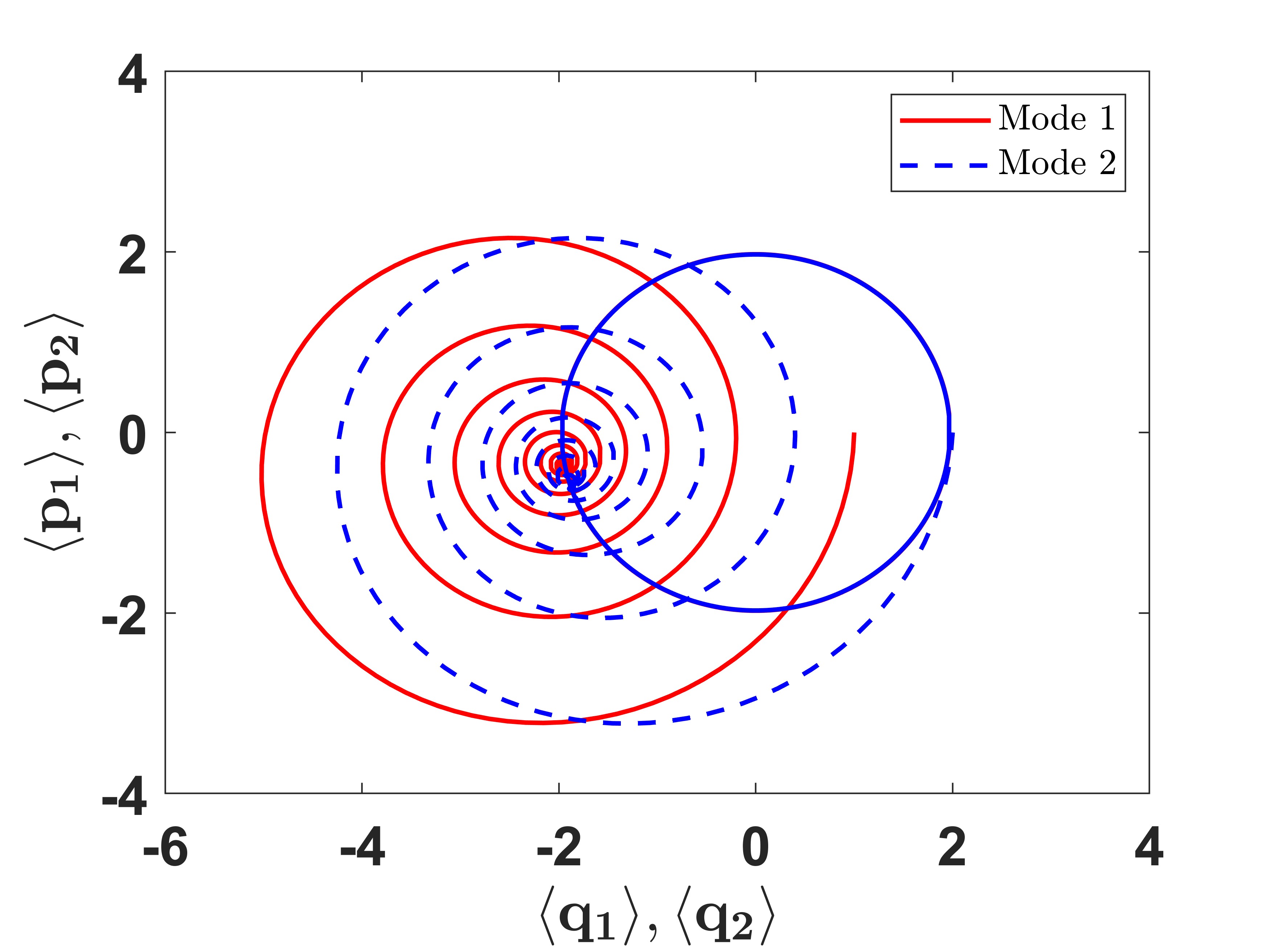}%
	 }
      \subfloat[\label{Q1Q2F}]{%
		\includegraphics[height=4.7cm,width=0.34\linewidth]{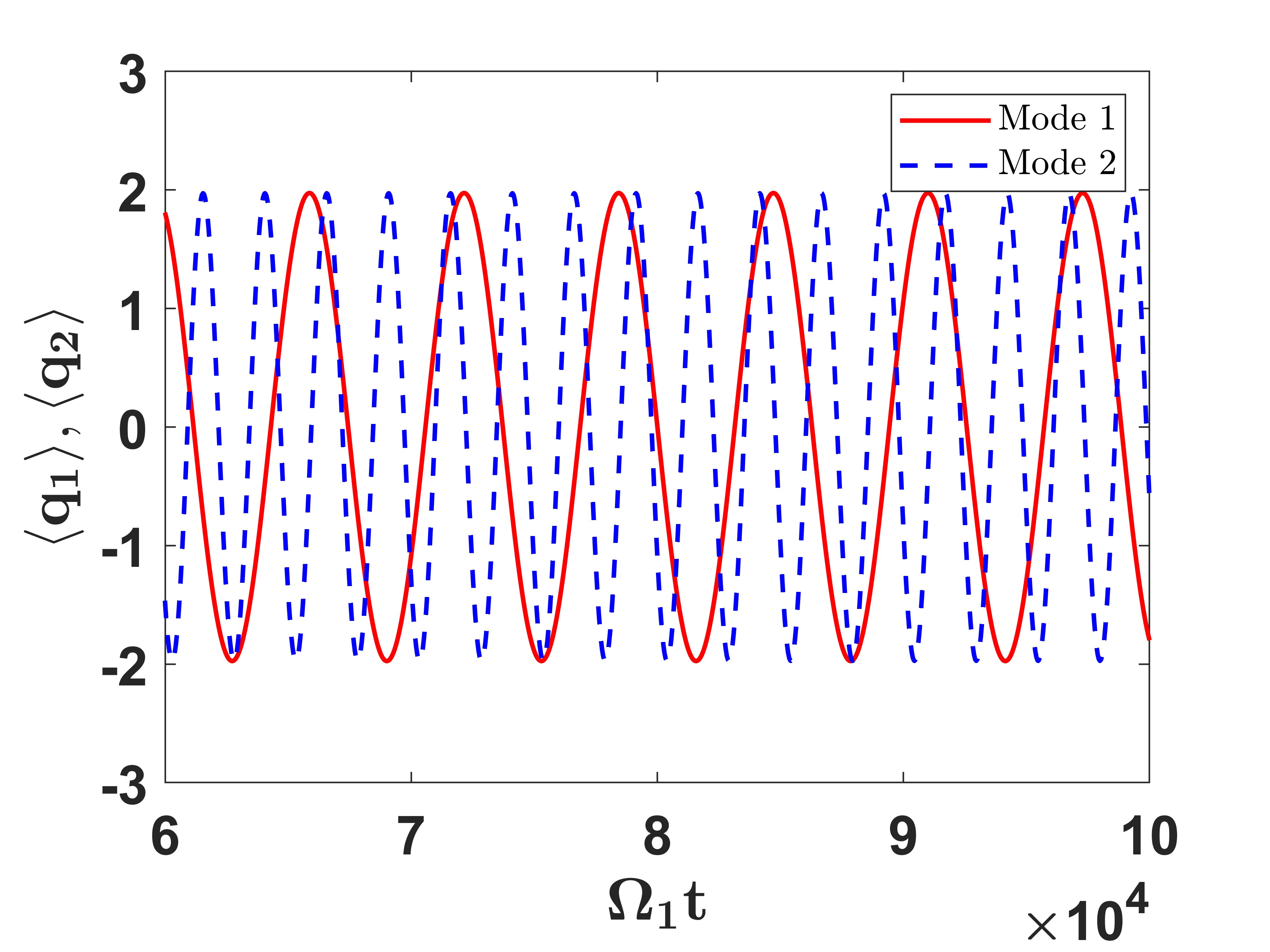}%
	 }
        \subfloat[\label{P1P2F}]{%
		\includegraphics[height=4.7cm,width=0.34\linewidth]{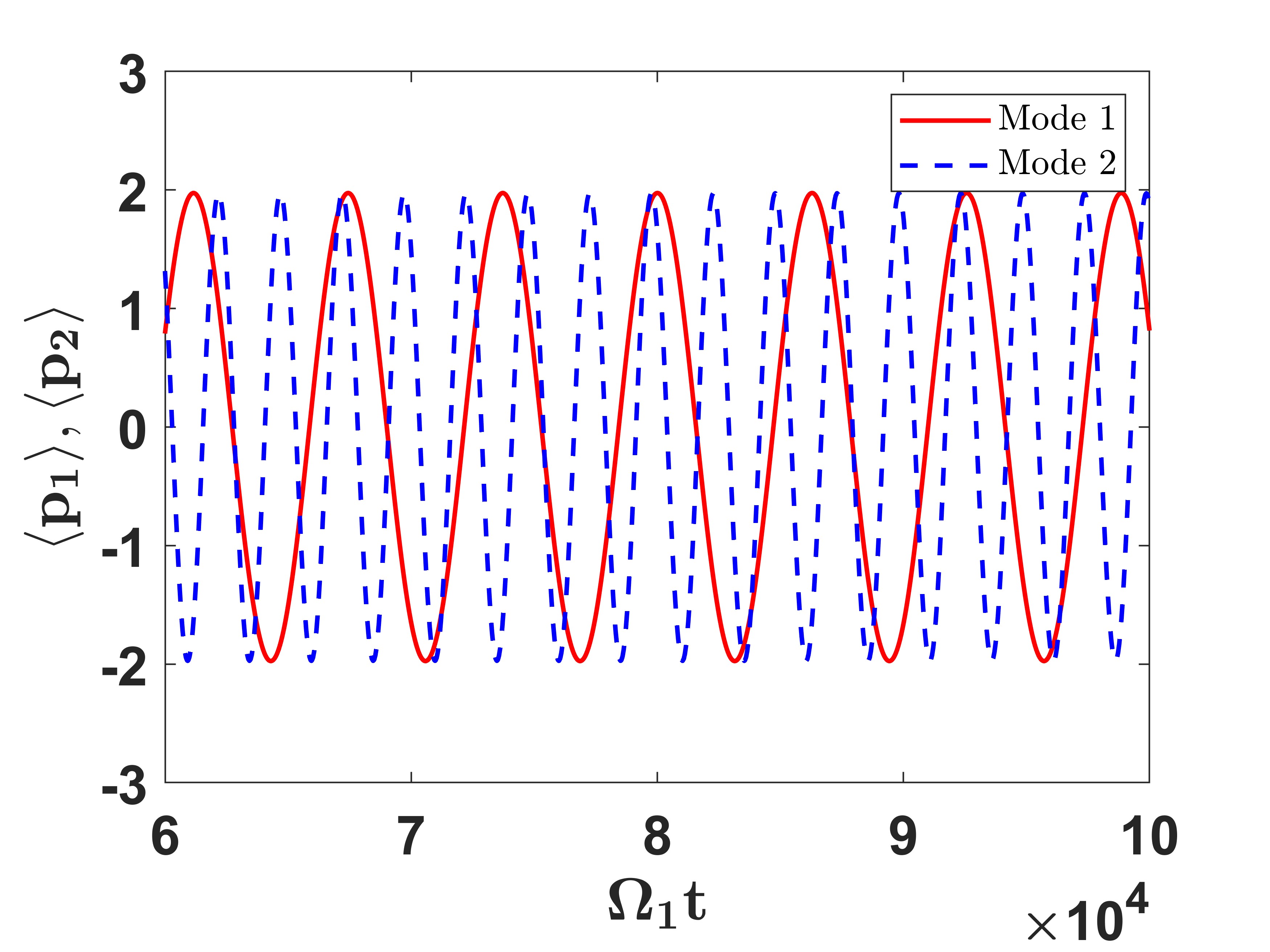}
	}\\
\caption{
Limit-cycle trajectories in the $\langle{q_{1}\rangle} \leftrightharpoons \langle{p_{1}\rangle}$ (red) and $\langle{q_{2}\rangle} \leftrightharpoons \langle{p_{2}\rangle}$ (blue) spaces (a, d, g); variation of the mean values $\langle{q}_{1}\rangle$ (red), $\langle{q}_{2}\rangle$ (blue) (b, e, h); and $\bar{p}_{1}$ (red), $\langle{p}_{2}\rangle$ (blue) (c, f, i). The parameters for each case are: (a--c) $\Omega_{1} = 1$, $\Omega_{2} = 1.00001$,$\Delta_{1} = \Delta_{2} = 0.001$. (d--f) $\Omega_{1} = 1$, $\Omega_{2} = 1.1$, $\Delta_{1} = \Delta_{2} = 0.001$. (g--i) $\Omega_{1} = 1$, $\Omega_{2} = 1.00001$,
$\Delta_{1} = 0.001$, $\Delta_{2} = 0.0025$. Other parameters (common to all cases): 
$\Omega_{c} = 1$, $\Delta_{c} = -0.2$, $g_{1} = g_{2} = 0.5$, 
$K_{1} = K_{2} = 10^{-10}$, and $\gamma_{1} = \gamma_{2} = \gamma_{c} = 0.1$. All parameters are normalized with respect to $\Omega_{1}$ with the initial conditions $(\langle{q}_1\rangle,\langle{p}_1\rangle)=(1,0)$ and $(\langle{q}_2\rangle,\langle{p}_2\rangle)=(2,0)$ [i.e., $\alpha_1=1/\sqrt{2}$ and $\alpha_2=\sqrt{2}$ respectively]. The limit cycle trajectories are shown till $\Omega_1t=10^5$.}
\label{Limitcycle}
\end{figure*}

\begin{figure*}[ht]
\centering
\includegraphics[height=4.5cm,width=6cm]{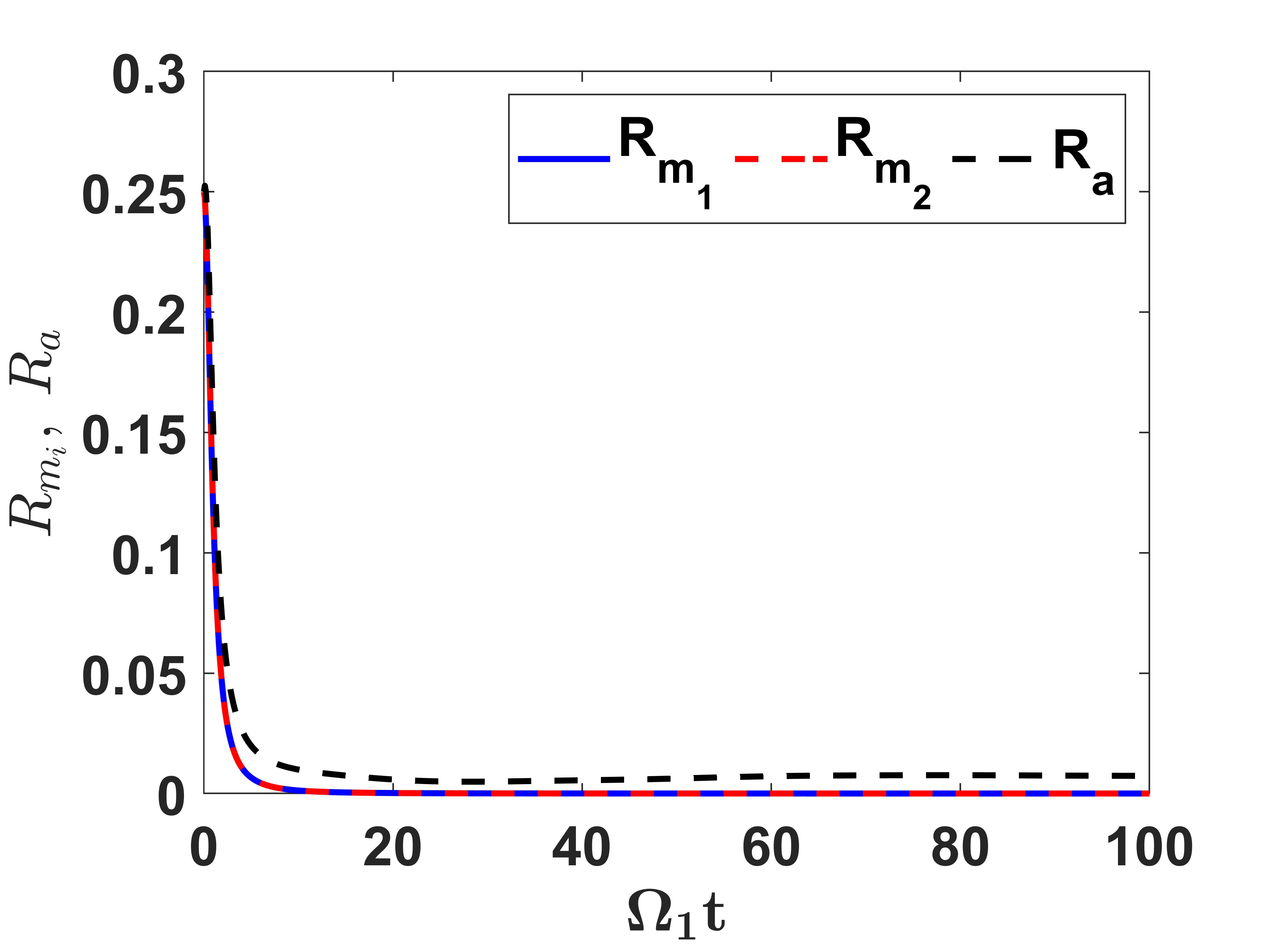}
\caption{Time evolution of the fluctuation-to-mean ratios 
$\langle \delta m_i^\dagger \delta m_i \rangle / |\alpha_i|^2$ ($i=1,2$) and 
$\langle \delta a^\dagger \delta a \rangle / |\beta|^2$, for the parameters as in Figs. \ref{limitcycle}.}
\label{fig:4}
\end{figure*}

\subsection{Solution in mean-field approximation}
Due to the analytical complexity of solving the nonlinear Eqs.~(\ref{b1eqn}) and~(\ref{b2eqn}), we employ the mean-field approximation to linearize these equations. In the limit of large excitation of the bosonic modes, the relevant operators can be expressed as the sum of their mean values and quantum fluctuations near the mean values, i.e., ${m}_i \rightarrow \langle{m}_i\rangle + \delta{m}_i = {\alpha}_i + \delta{m}_i$ and ${a} \rightarrow \langle{a}\rangle + \delta{a} = {\beta} + \delta{a}$. The approximation requires that relative fluctuations remain small, i.e., $|\alpha_i|^2 \gg \langle \delta m_i^\dagger \delta m_i \rangle$, $|\beta|^2 \gg \langle \delta a^\dagger \delta a \rangle$, $|\langle \delta m_i\rangle| / \alpha_i,\ |\langle \delta a\rangle| / \beta  \ll 1$, so that higher-order fluctuation terms can be neglected. The equations for these mean values $\alpha_i$ and $\beta$ are written as, 
\begin{equation}\label{alpha_ieqn}
\begin{aligned}
\dot{\alpha}_i &= -\frac{\gamma_i}{2} \alpha_i + \iota \bigg[ g_i \beta^* A_i e^{\iota t (B_i - \langle C_i\rangle)}\alpha_i^2 - \Omega_i e^{\iota t (D_i + \langle C_i\rangle)}  \\
&\quad - \Omega_i \alpha_i^* A_i e^{\iota t (D_i + \langle C_i\rangle)} \alpha_i - g_i e^{-\iota t (B_i - \langle C_i\rangle)} \beta \\
&\quad- g_i |\alpha_i|^2\beta A_i e^{-\iota t (B_i - \langle C_i\rangle)} + \Omega_i A_i e^{-\iota t (D_i + \langle C_i\rangle)} \alpha_i^2 \bigg]\;,
\end{aligned}
\end{equation}

\begin{equation}\label{beta eqn}
\begin{aligned}
\dot{\beta} &= -\frac{\gamma_c}{2} \beta - \iota \left( g_1 e^{\iota t (B_1 - \langle C_1 \rangle)} \alpha_1 + g_2 e^{\iota t (B_2 - \langle C_2\rangle)} \alpha_2 \right) - \iota \Omega_c e^{\iota t \Delta_c}\;.
\end{aligned}
\end{equation}

To represent the magnon dynamics in phase space, we introduce the quadrature operators 
$q_i=\frac{1}{\sqrt{2}}(m_i^{\dagger}+m_i)$ and 
$p_i=\frac{\iota}{\sqrt{2}}(m_i^{\dagger}-m_i)$ for the $i$th magnon mode, 
and for the cavity mode 
$x=\frac{1}{\sqrt{2}}(a^\dagger+a)$ and 
$y=\frac{\iota}{\sqrt{2}}(a^\dagger-a)$. 
These operators define the canonical phase-space variables of the corresponding cavity modes. The expectation values of the quadratures, $\langle q_i\rangle$ and $\langle p_i\rangle$, determine the real and imaginary parts of $\alpha_i$, respectively, such that 
$\alpha_i=\left(\langle q_i\rangle+\iota\langle p_i\rangle\right)/\sqrt{2}$. 
Analogously, the cavity amplitude is expressed in terms of its phase-space coordinates as 
$\beta=\left(\langle x\rangle+\iota\langle y\rangle\right)/\sqrt{2}$. 
Thus, the complex amplitudes $\alpha_i$ and $\beta$ are uniquely specified by the corresponding quadrature expectation values, providing a direct geometrical interpretation of the mean-field dynamics in phase space.

In the case when the quantum fluctuations can be ignored, the classical synchronization prevails and may be quantified as 
\begin{equation}
S_{c}(t) = \left\langle q_{-}^{2}(t) + p_{-}^{2}(t) \right\rangle^{-1} \;,\label{eq:sc}
\end{equation}
where
\begin{eqnarray}
q_{-}(t)& =& \frac{1}{\sqrt{2}}\left[ q_{1}(t) - q_{2}(t) \right]\;,\nonumber\\
p_{-}(t) &=& \frac{1}{\sqrt{2}}\left[ p_{1}(t) - p_{2}(t) \right]\;
\end{eqnarray}
are the differences between the position quadratures $q_{1,2}$ of the two oscillators and between their linear momentum quadratures $p_{1,2}$, respectively. In the limit, when $q_-,p_-\rightarrow 0$, the $S_c \rightarrow \infty$ denotes complete classical synchronization. In phase space, they would represent a limit cycle, with a constant phase difference between them.

\subsection{Quantum regime}

To describe the dynamics beyond the semiclassical approximation, we retain the quantum fluctuations around the steady-state mean values and neglect higher-order nonlinear terms in the fluctuation operators. Therefore, the equations of motion for the fluctuation operators take the form
\begin{eqnarray} \label{c1}
\delta\dot{m}_i &=& P_i\delta{m}_i + Q_i\delta{m}_i^{\dagger}+ R_i\delta{a} + S_i\delta{a}^{\dagger}+ \sqrt{\gamma_i} m_{\text{in}}^{(i)} \;, \nonumber \\
\delta\dot{a} &=&\sum_{i=1}^2 \left(U_i\delta{m}_i+W_i\delta{m}_i^{\dagger}\right)+T\delta{a} + \sqrt{\gamma_c} a_{\text{in}} \;,
\end{eqnarray}    
The expressions of the parameters $P_i$, $Q_i$, $R_i$, $S_i$, $U_i$, $W_i$, and $T$ are listed in Appendix A.

To ascertain the degree of quantum synchronization between oscillators, we adopt a figure of merit originally proposed by Mari {\it et al.} \cite{PhysRevLett.111.103605}.
In this regime, when quantum fluctuations cannot be ignored, the quadrature variables need to be redefined with respect to their expectation values as follows:
\begin{eqnarray}
q_{-}(t) &\rightarrow q_{-}(t) - \langle q_-(t)\rangle = \delta q_{-}(t)\;,\nonumber\\
p_{-}(t) &\rightarrow p_{-}(t) - \langle p_-(t)\rangle = \delta p_{-}(t)\;.
\end{eqnarray}  
The quantum synchronization measure can then be expressed only in terms of these fluctuations $\delta q_- = \delta q_1-\delta q_2$ and $\delta p_-=\delta p_1-\delta p_2$ as 
\begin{equation}
S_{q}(t) \equiv \left\langle \delta q_{-}^{2}(t) + \delta p_{-}^{2}(t) \right\rangle^{-1} \;.\label{eq:sq}
\end{equation}
Here, we define the dimensionless quadrature fluctuations as $\delta q_i = \frac{1}{\sqrt{2}}\left( \delta m_i^\dagger + \delta m_i \right)$ and $
\delta p_i = \frac{\iota}{\sqrt{2}}\left( \delta m_i^\dagger - \delta m_i \right)$.

In case of a constant phase difference $\phi = \phi_{2} - \phi_{1}$ between the limit cycles, the generalized form of the quadrature differences can be written as $q_{-}^\phi(t)=\frac{1}{\sqrt{2}}\left[q_1^\phi(t)-q_2^\phi(t)\right]$ and $p_{-}^\phi(t)=\frac{1}{\sqrt{2}}\left[p_1^\phi(t)-p_2^\phi(t)\right]$, where
\begin{eqnarray}
q_i^\phi(t) & =q_i(t) \cos \left(\phi_i\right)+p_i(t) \sin \left(\phi_i\right),\nonumber\\
p_i^\phi(t) & =p_i(t) \cos \left(\phi_i\right)-q_i(t) \sin \left(\phi_i\right)\;,
\end{eqnarray}
and $\phi_{i} = \tan^{-1}\left[\frac{\langle{p}_{i}\rangle(t)}{\langle{q}_{i}\rangle(t)}\right]$.
Therefore, the generalized form of the Eq.~(\ref{eq:sq}) - the so-called ``quantum $\phi$ synchronization" \cite{qiao2020quantum, sun2024quantum} - can be defined as 
\begin{equation}
S_{q}^{\phi}(t) = \left\langle \delta q_{-}^{\phi}(t)^{2} + \delta p_{-}^{\phi}(t)^{2} \right\rangle^{-1} \label{eq:sqm}\;.
\end{equation}
Note that the above quantifier of synchronization is a marker of classical phase-locked complete quantum synchronization, and not  `quantum phase synchronization'. Within this definition, $S_q^{\phi}(t)$ is invariant under arbitrary choices of the constant phase offset $\phi$, and therefore depends only on the relative phase locking between the two limit cycles rather than on the absolute phase reference. 

To calculate $S_{q}^{\phi}(t)$, we need to solve for the quadrature fluctuations of the oscillators. Therefore, it becomes more convenient to obtain the corresponding equations from Eqs.~(\ref{c1}), by replacing the relevant fluctuation operators and input noise operators with their quadratures, namely, $\delta x = \frac{1}{\sqrt{2}} (\delta a^\dagger + \delta a)$, $\delta y = \frac{\iota }{\sqrt{2}} (\delta a^\dagger - \delta a)$, $ q_{\rm in}=\frac{1}{\sqrt{2}}\left( m_{\rm in}^{\dagger}+ m_{\rm in}\right)$, $ p_{\rm in}=\frac{\iota}{\sqrt{2}}\left( m_{\rm in}^{\dagger}- m_{\rm in}\right)$, $ x_{\rm in} = \frac{1}{\sqrt{2}} ( a_{\rm in}^\dagger + a_{\rm in})$, and $ y_{\rm in} = \frac{\iota}{\sqrt{2}} (a_{\rm in}^\dagger -a_{\rm in})$. 
The equations for quadrature fluctuations then take a consolidated form as given by 
\begin{equation}
 \dot{Y}(t) = M(t) Y(t) + N(t) \;,
\end{equation}
where $Y(t)^\intercal=(\delta q_1,\delta p_1, \delta q_2, \delta p_2, \delta x, \delta y)$, 
$M(t) = [M_{ij}]_{6\times6}$
is a time-dependent coefficient matrix, and $\intercal$ denotes matrix transposition.

\begin{figure*}

      \subfloat[\label{Sq_1}]{%
		\includegraphics[height=4.5cm,width=0.36\linewidth]{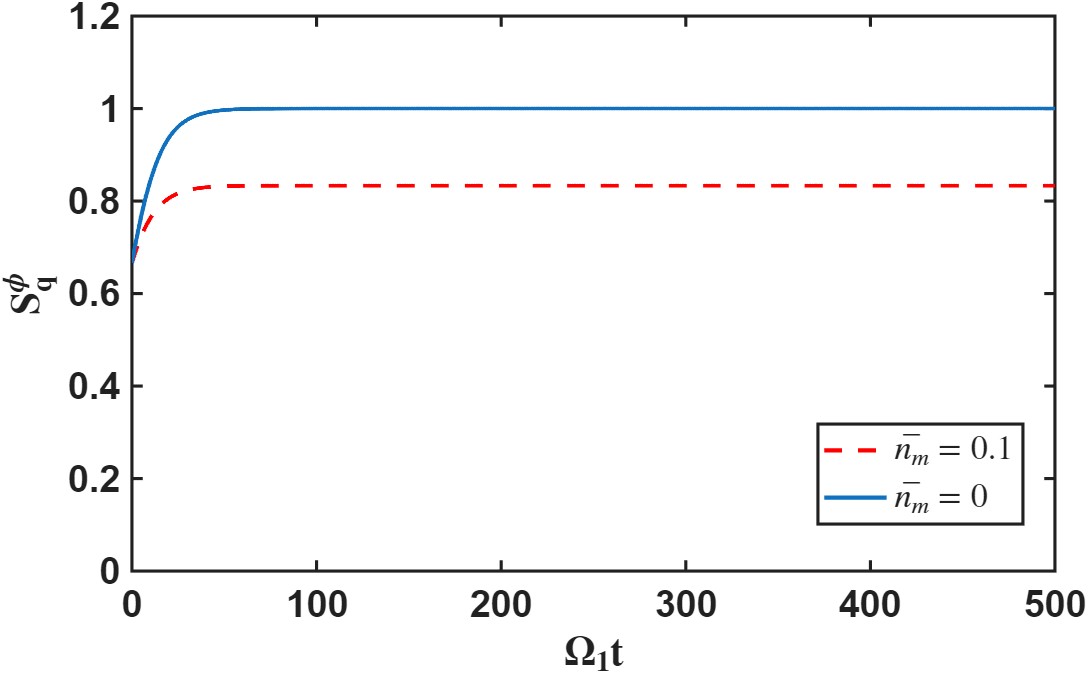}
	}
    \subfloat[\label{synchronization temp}]{%
		\includegraphics[height=4.5cm,width=0.36\linewidth]{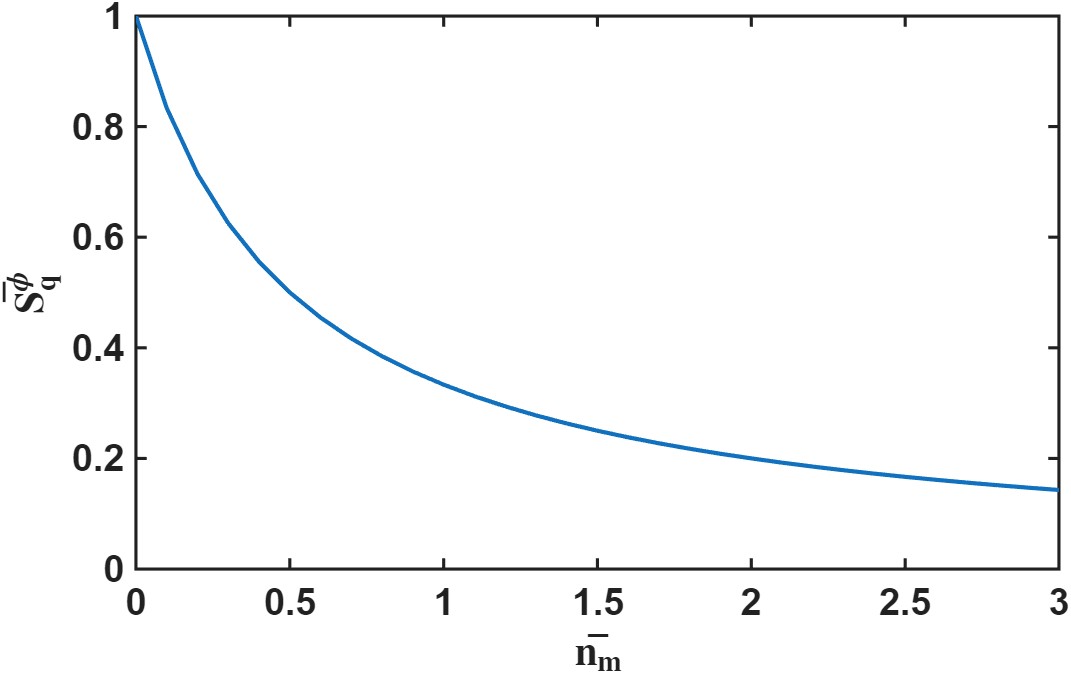 }
	}

\caption{(a) Variation of $S_{q}^{\phi}$, with respect to time $t$ for $\bar{n}_m =0$ (blue solid) and $\bar{n}_m = 0.1$ (red dashed), (b) variation of $\bar{S}_{q}^{\phi}$, with respect to the mean phonon number $\bar{n}_{m}$ of the environment. The other parameters are
 $g_1=g_2=0.1$, $K_1=K_2=10^{-10}$, $ \Omega_{1}=1$, $\Omega_{2}=1.1$,
$\Omega_{c}=1$,
$\Delta_1 = \Delta_2 =0.001$, $\Delta_c =-0.2$, $\gamma_1=\gamma_2=\gamma_c=0.1$, and $\phi= 0.1320$ radian. We have chosen an initial condition $(\langle{q}_1\rangle,\langle{p}_1\rangle) = (1,0)=(\langle{q}_2\rangle,\langle{p}_2\rangle)$ [i.e.,$\alpha_i=1/\sqrt{2}$].}
\label{Synchronization}
\end{figure*}

The vector $N(t)$ containing the noise terms is given below:
\begin{eqnarray}
N(t)^\intercal&=&\left( \sqrt{\gamma_1}\delta q_{\text {in}}^{(1)}, \sqrt{\gamma_1}\delta p_{\text {in}}^{(1)},\sqrt{\gamma_2}\delta q_{\text {in}}^{(2)} ,\sqrt{\gamma_2}\delta p_{\text {in}}^{(2)},\right.\nonumber\\
&&\left.\sqrt{\gamma_c}\delta x_{\text {in}}, \sqrt{\gamma_c}\delta y_{\text {in}}\right)\;.
\end{eqnarray}

The Eq.~(\ref{eq:sqm}) involves correlations among different quadratures at a given time $t$. Therefore, it is important to get the time-dependent solutions of these correlations, instead of the solution of the quadrature itself.  This can be done in terms of the following linear time-dependent equation of the matrix $C$ of covariances:
\begin{equation}\label{corr}
\dot{C}(t)=M(t)C(t) + C(t) M(t)^\intercal+D\;,
\end{equation}

and

\begin{equation}
M(t)=
\begin{pmatrix}
P_1 & Q_1 & 0 & 0 & R_1 & S_1 \\
Q_1^* & P_1^* & 0 & 0 &  S_1^* & R_1^* \\
0 & 0 & P_2 & Q_2 &  R_2 & S_2  \\
0 & 0 & Q_2^* & P_2^* & S_2^* & R_2^*\\
U_1 & W_1 & U_2 & W_2 & T & 0 \\
W_1^* & U_1^* & W_2^* & U_2^* & 0 & T^*
\end{pmatrix},
\end{equation}
where the elements of $C$, $C_{ij}= \left[\left\langle Y_i(t)Y_j(t)+Y_j(t)Y_i(t)\right\rangle \right]/2 $, represent the correlation between two elements $Y_{i,j}$ at a time $t$. The diffusion matrix $D$ is given by
$D = {\rm diag}[ V_1, V_1, V_2, V_2, V_3, V_3]$
where $V_i=\gamma_i\left( \bar{n}_{m} + 0.5 \right)$ ($i\in 1,2$), and $V_3=\gamma_c\left( \bar{n}_{m} + 0.5 \right)$.
In the matrix \( C \), each block along the diagonal represents the \( 2 \times 2 \) covariance matrix corresponding to an individual mode, while each off-diagonal block \( C_{ij} \) denotes the \( 2 \times 2 \) inter-mode covariance between modes \( i \) and \( j \). 

We solve the Eq.~(\ref{corr}) using the initial condition $C(0)={\rm diag}[0.75,0.75,0.75,0.75,0.75,0.75]= 0.75\;\mathbb{1}_{6\times6}$, which is a positive definite matrix and satisfies the uncertainty principle \cite{simon2000peres}. The complete quantum synchronization $S_{q}^\phi(t)$ can then be expressed in a concise form as
\begin{eqnarray}\label{sync}
S_{q}^\phi(t) &= &2 \left[C_{11}(t) + C_{22}(t) + C_{33}(t) + C_{44}(t) \right.\nonumber\\
&&+ 2 \sin{\phi}\{C_{23}(t) 
- C_{14}(t)\} \nonumber\\
&&\left.- 2\cos{\phi}\{C_{13}(t) + C_{24}(t)\}\right]^{-1}\;,
\end{eqnarray}
which lies between 0 and 1. Here, $S_{q}^\phi(t) =1$ indicates complete synchronization, and a null $S_{q}^\phi(t)$ denotes no synchronization.

\section{Results}

In this section, we will discuss by numerically solving the Eqs.~(\ref{alpha_ieqn}), (\ref{beta eqn}), and (\ref{corr}), how the quantum synchronization between two magnon modes is developed. We consider a strong driving regime, when the Rabi frequencies $\Omega_i$ and $\Omega_c$ of the driving fields are much larger than the decay rates $\gamma_i$ of the $i$th magnon mode and $\gamma_c$ of the cavity mode, respectively. This regime would otherwise render an independent dynamics of the magnon mode in the absence of the cavity mode, exhibiting an underdamped oscillation that could sustain up to a time scale $t\sim 1/\gamma_i$.  In the present case, the cavity mode induces an indirect coupling between the magnon modes, leading to classical synchronization between them that persists on a timescale much longer than $1/\gamma_i$. We show these results in Figs.~\ref{limitcycle}. 

We first choose  $\Delta \Omega=\Omega_2 -
\Omega_1= 10^{-5}\Omega_1$, for which both the magnon modes rapidly converge to a shared limit cycle in phase space (Figs.~\ref{pqa},~\ref{q1q2a}, and~\ref{p1p2a}). This behavior is a clear signature of classical synchronization, where the mean quadratures $\langle{q}_1\rangle$, $\langle{q}_2\rangle$ and $\langle{p}_1\rangle$, $\langle{p}_2\rangle$ become indistinguishable at times as long as $t\approx 10^5/\Omega_i \gg 1/\gamma_i$. Such phase-locked dynamics of two YIG spheres have been previously explored in a cavity-magnomechanical setup, in which the phase synchronization has been achieved between the vibrational degrees of freedom, instead of magnon modes \cite{PhysRevResearch.5.043197}. When the Rabi frequency $\Omega_2$ is increased so that $\Delta\Omega= 0.1\Omega_1$, the system still maintains the phase synchronization, with a nonzero phase difference $\phi = 0.1320$ radian (Figs.~\ref{pqb},~\ref{q1q2b}, and~\ref{p1p2b}) between the trajectories of the two magnon modes. However, the quadratures still oscillate at the same frequency. Such a behavior refers to a classical phase synchronization.

We further consider that the coupling strengths $g_i$ of these magnon modes with the cavity are different from each other. This could arise when the YIG spheres are trapped inside the cavity in an asymmetrical configuration with respect to the central antinodes of the cavity. When we choose $g_1>g_2$, the amplitude of the second magnon mode becomes lower compared to that of the first. Nevertheless, the system maintains phase synchronization with a phase difference of $\phi = 1.3357$ radian (Fig. \ref{PQC}, \ref{Q1Q2C} and \ref{P1P2C}). We note that such a phase relationship between coupled magnonic oscillators can be continuously tuned by varying coupling strengths \cite{PhysRevResearch.5.043197, PhysRevResearch.6.033207}.

One of the important features of classical synchronization is its robustness against a large change in initial conditions. We verify the existence of limit cycles by varying the initial conditions (Fig. \ref{PQD}, \ref{Q1Q2D}, and \ref{P1P2D}). We choose an initial condition $[\langle{q}_1\rangle(0),\langle{p}_1\rangle(0)]=[1,0]$ and $[\langle{q}_2\rangle(0),\langle{p}_2\rangle(0)]=[2,0]$. For $\Delta \Omega = 10^{-5}\Omega_1$, both magnon modes rapidly converge to a common limit cycle in phase space. Similar feature persists also when the Rabi frequency $\Omega_2$ is increased so that $\Delta \Omega = 0.1\Omega_1$, though with a finite phase difference of $\phi = 1.2276$ radian (Figs. \ref{PQE}, \ref{Q1Q2E}, and \ref{P1P2E}). In this regime, the system exhibits chaotic transients before settling into a shared limit cycle, providing a clear signature of synchronization. Furthermore, when the detunings of the driving fields are chosen differently, e.g., $\Delta_1 = 0.001$ and $\Delta_2=0.0025$, the system continues to display phase synchronization with an increased phase difference of $\phi = 1.5337$ radian (Figs.~\ref{PQF}, \ref{Q1Q2F}, and \ref{P1P2F}).

We have further examined the validity of the semiclassical (mean-field) approximation
by quantitatively comparing the coherent mean occupations with the corresponding quantum fluctuation contributions. Specifically, we have evaluated the ratios
$
R_{m_i} = \langle \delta m_i^\dagger \delta m_i \rangle/|\alpha_i|^2,
R_{a} = \langle \delta a^\dagger \delta a \rangle/|\beta|^2
$ and display their time-dependence in Fig. \ref{fig:4} for the same parameters as used in Fig. \ref{limitcycle}. Note that the second-order moments of the fluctuation operators can be written as linear combinations of the elements of the covariance matrix
$C$, as $
\langle \delta m_1^\dagger \delta m_1 \rangle= \frac{1}{2}\left( C_{11} + C_{22} - 1 \right)$,
$\langle \delta m_2^\dagger \delta m_2 \rangle= \frac{1}{2}\left( C_{33} + C_{44} - 1 \right), $
and $\langle \delta a^\dagger \delta a \rangle= \frac{1}{2}\left( C_{55} + C_{66} - 1 \right)$. Clearly, at the long time-scale, when the quantum synchronization sets in (as we show next), these ratios remain much smaller than unity, e.g., $R_{m_1} = R_{m_2}  = 2.4769\times 10^{-7}$, and $R_a = 7.46\times 10^{-3}$, thereby justifying the mean-field approximation.

To quantify quantum synchronization, we next compute the $S_{q}^{\phi}$. Our results show that $S_{q}^{\phi}$ stabilizes at unity ($S_{q}^{\phi} = 1$) at long times, indicating complete quantum synchronization between the magnon modes (see Fig. \ref{Sq_1}). We have observed that the optimal value of $S_{q}^{\phi}$ remains unity for a wide range of the relative difference of the coupling constant $\Delta\Omega/\Omega_1$ and of the detunings $(\Delta_2-\Delta_1)$. We have verified that $S_q^\phi$ remains unity at long times, even when the two magnon modes start from different initial conditions (results not shown). It is interesting to note that the $S_q^\phi$ becomes unity at a time scale $t\sim 50/\Omega_1$, which is much larger than the decay time-scale $1/\gamma_1 \sim 10/\Omega_1$.

In contrast, for a small thermal population $\bar n_m=0.1$, $S_q^{\phi}$ still grows monotonically but saturates at a reduced value $S_q^{\phi}\simeq 0.83$, reflecting partial degradation of synchronization due to thermal noise. Nevertheless, in both cases the system reaches a steady synchronized regime, demonstrating that cavity-mediated magnon synchronization remains robust against weak thermal decoherence.

To examine the impact of decoherence, we plot the time-averaged synchronization $\bar{S}_q^{\phi}$ as a function of the mean thermal phonon number $\bar{n}_m$ (see Fig.~\ref{synchronization temp}). As $\bar{n}_m$ increases, quantum synchronization degrades (see \cite{ harraf2025quantum, amazioug2023feedback, PhysRevLett.122.187701}, for similar results). Note that such a degradation can be controlled by using magnon squeezing and photon tunneling, as shown in \cite{harraf2025quantum, amazioug2023feedback}. Notably, Harraf \textit{et al.}~\cite{harraf2025quantum} reported that quantum synchronization is more robust to thermal effects than entanglement is. 

\subsection{Experimental consideration}
Our parameter choices reflect realistic regimes established in cavity magnonics and magnomechanics, readily available to demonstrate our results. Normalized frequencies ($\Omega_{1,2}$,$\Omega_c\sim 1$) correspond to 1 GHz, typical for YIG-based systems \cite{zhang2016cavity,li2019entangling}. The small detuning ($\Delta\Omega= 10^{-5}\Omega_1$) is achievable via local magnetic field tuning \cite{li2019entangling}. Coupling strengths ($g_{1,2} = 0.1$) represent the strong-coupling regime $g_1\sim \gamma_i,\gamma_c$ (10–100 MHz), while damping rates ($\gamma_{1,2,c} = 0.1$) match high-quality YIG spheres and microwave cavities \cite{zhang2016cavity}. Kerr nonlinearities ($K_{1,2} = 10^{-10}$) refer to weak intrinsic nonlinearity, consistent with low excitation levels \cite{wang2016magnon}. These parameters enable robust synchronization and align with recent experimental and theoretical studies~\cite{PhysRevResearch.5.043197,li2019entangling}, grounded in well-established coupling mechanisms \cite{kittel1958interaction}.

\section{Conclusion} 
In this work, we presented a comprehensive theoretical framework to investigate the synchronization between two magnon modes mediated by a single-mode microwave cavity. The magnon modes do not interact directly but are indirectly coupled through their mutual interaction with the cavity mode, which induces an effective nonlinear coupling. This nonlinear interaction, combined with intrinsic dissipation in the system, gives rise to complex dynamical behavior.
Our analysis reveals that magnon modes can exhibit two distinct forms of synchronization. First, we observed classical synchronization with different initial conditions (Fig.~\ref{limitcycle}) and (Fig.~\ref{Limitcycle}), where the dynamical trajectories of the magnon modes become phase-locked and evolve coherently and regularly over time. 

Second, and more intriguingly, we observed the emergence of quantum synchronization between the coupled magnon modes (Fig.~\ref{Synchronization}). This phenomenon is quantitatively characterized by the synchronization measure \(S_q^{\phi}\), which approaches unity under optimal conditions. In this regime, the individual quantum fluctuations of each mode become phase-locked, leading to coherent evolution despite the presence of decoherence and thermal noise. This indicates that the system dynamically suppresses relative quantum uncertainties, allowing for robust synchronization at the quantum level. In addition, we explored the effect of temperature on this delicate quantum behavior (Fig.~\ref{Synchronization}). We find that as thermal noise increases, the degree of quantum synchronization gradually decreases. This finding underscores the importance of low-temperature environments when trying to observe and maintain quantum synchronization in experiments.

In summary, our study provides clear evidence that both classical and quantum synchronization are achievable in cavity-coupled magnon systems, and it highlights the critical role of system parameters and thermal noise in shaping this behavior. This understanding could have future applications in areas like quantum communication and computation, where controlling synchronization at the quantum level is essential.

\section{ACKNOWLEDGMENTS}

One of us (J. Ghildiyal) acknowledges the financial support provided by the Department of Science and Technology-Innovation in Science Pursuit for Inspired Research (DST-INSPIRE), Government of India, through the fellowship DST/INSPIRE Fellowship/2019/IF190615 during this work.

\appendix

\section{Relevant coefficients in Eqs. (\ref{c1})}

{\footnotesize
\begin{align*}
P_i &= -\frac{\gamma_i}{2} 
      + \iota \Bigg[
        A_i g_i \alpha_i \beta^*\!\left(2-A_i |\alpha_i |^2\right)e^{\iota t(B_i-\langle C_i \rangle)}
        -A_i\Omega_i \alpha_i^* e^{\iota  t(D_i+\langle C_i \rangle)} \\
    &\qquad -g_iA_i \alpha_i^* \beta e^{-\iota t(B_i-\langle C_i \rangle)}
        -A_i\Omega_i \alpha_i^*e^{\iota t(D_i+\langle C_i \rangle)}\left(1+A_i |\alpha_i|^2 \right) \\
    &\qquad +A_i\Omega_i \alpha_i e^{-\iota t(D_i+\langle C_i \rangle)}\left(2-A_i |\alpha_i|^2 \right) \\
    &\qquad -A_i g_i \alpha_i^* \beta e^{-\iota t(B_i-\langle C_i \rangle)}\left(1+A_i |\alpha_i|^2 \right)
      \Bigg] \;,
\end{align*}

{\footnotesize
\begin{align*}
Q_i &=  \iota \Bigg[
        -A_i^{2} g_i \alpha_i^{3} \beta^*
        -A_i\Omega_i \alpha_i e^{\iota t(D_i+\langle C_i \rangle)}
        -A_i g_i \alpha_i \beta e^{-\iota t(B_i-\langle C_i \rangle)} \\
    &\qquad -A_i\Omega_i \alpha_i\left(1+A_i |\alpha_i|^2 \right)e^{\iota t(D_i+\langle C_i \rangle)}
        -A_i^{2}\Omega_i \alpha_i^{3} e^{-\iota t(D_i+\langle C_i \rangle)} \\
    &\qquad -A_i g_i \alpha_i \beta \left(1+A_i |\alpha_i|^2\right)e^{-\iota t(B_i-\langle C_i \rangle)}
      \Bigg]\;,
\end{align*}
}

{\footnotesize
\begin{align*}
R_i &= -\iota g_i\left(1 +A_i |\alpha_i|^2 \right)e^{-\iota t(B_i-\langle C_i \rangle)}\;,
\end{align*}
}

{\footnotesize
\begin{align*}
S_i &= \iota g_i A_i \alpha_i^{2} e^{\iota t(B_i-\langle C_i \rangle)}\;,
\end{align*}
}

{\footnotesize
\begin{align*}
U_i &= - \iota g_i \left(1 - A_i |\alpha_i|^2\right)e^{\iota t(B_i - \langle C_i \rangle)}\;,
\end{align*}
}

{\footnotesize
\begin{align*}
W_i &= \iota A_i g_i \alpha_i^{2} e^{\iota t(B_i - \langle C_i \rangle)}\;,
\end{align*}
}

{\footnotesize
\begin{align*}
T &= -\dfrac{\gamma_c}{2}\;,
\end{align*}
}

where $i\in 1,2$.

\bibliographystyle{ieeetr}

\bibliography{main}

\end{document}